\documentstyle[epsf]{mn}
\begin{document}
\title[H{\normalsize \it I} observations of low surface brightness
galaxies]{H{\Large\bf I} observations of low surface brightness
galaxies: Probing low-density galaxies}
\author[W.J.G. de Blok, S.S. McGaugh and J.M. van der Hulst]
{W.J.G.~de~Blok,$^1$ S.S. McGaugh$^2$\thanks{Present address: Department
of Terrestrial Magnetism, Carnegie Institution of Washington,
5241 Broad Branch Road NW, Washington, DC 20015, USA}
and J.M.~van~der~Hulst$^1$\\
$^{1}$Kapteyn Astronomical Institute, P.O.~Box 800, 9700 AV
Groningen, The Netherlands\\
$^{2}$Institute of Astronomy, University of Cambridge,  Madingley Road,
Cambridge CB3 0HA, UK}
\date{received: ; accepted: }
\maketitle
\begin{abstract}
 We present Very Large Array ({\sc vla}) and Westerbork Synthesis Radio
Telescope ({\sc wsrt}) 21-cm H{\sc i} observations of 19
late-type low
surface brightness (LSB) galaxies.  Our main findings are that these 
galaxies, as well as having low surface brightnesses, have low
H{\sc i} surface densities, about a factor of $\sim 3$ lower
than in normal late-type galaxies. We
show that LSB galaxies in some respects
resemble the outer parts of late-type normal galaxies, but may be less
evolved.  LSB galaxies are more gas-rich than their high surface
brightness counterparts.  The rotation curves of LSB galaxies rise more
slowly than those of HSB galaxies of the same luminosity, with
amplitudes between 50 and 120~km~s$^{-1}$, and are often still
increasing at the outermost measured point.  The shape of the rotation
curves suggests that LSB galaxies have low matter surface
densities.  We use the average total mass surface density 
of a galaxy as a measure for the
evolutionary state, and show that LSB galaxies are among the least
compact, least evolved galaxies.  We show that both $M_{\rm HI}/L_B$ and
$M_{\rm dyn}/L_B$ depend strongly on central surface brightness, consistent
with the surface brightness--mass-to-light ratio relation required by
the Tully-Fisher relation.  LSB galaxies are therefore
slowly evolving galaxies, and may well be low surface density systems in
all respects. 
 \end{abstract}
\section{Introduction}

In recent years low surface brightness (LSB) galaxies have received an
increasing amount of attention.  With the advent of deep, large-field
surveys many of these galaxies have been discovered, both in clusters
and in the field (Schombert et al.  1988, 1992 [hereafter SBSM], Turner
et al.  1993, Davies et al.  1994, Schwartzenberg et al. 
1995, Sprayberry et al.  1995a). 

The main property which distinguishes LSB galaxies from normal
``classical'' galaxies that define the Hubble sequence is their low
surface brightness ($\mu_0(B) \ga 23$) and not their size.  LSB galaxies
are not necessarily dwarfs.  Many of them are disk galaxies that
typically have scale lengths of several kpc, (McGaugh \& Bothun 1994, de
Blok et al.  1995 [hereafter BHB95]) i.e., in the same size range as typical high surface
brightness (HSB) galaxies.  The faintest LSB disk galaxies hitherto
discovered have central surface brightnesses of $\sim 26$ mag
arcsec$^{-2}$ (Turner et al.  1993; Irwin et al.  1990; Davies, Phillips
\& Disney 1988).  This is almost two orders of magnitude fainter than
those of the disk galaxies that obey Freeman's Law (1970), which states
that most disk galaxies have a central $B$-band surface brightness in
the range $21.65 \pm 0.30$ mag arcsec$^{-2}$. 

The existence of LSB galaxies with central surface brightnesses
deviating many $\sigma$ from the Freeman value indicates that this law
is not a complete description of the central surface brightness
distribution of disk galaxies.  Attempts to construct the proper
distribution from extant data (McGaugh, Bothun \& Schombert 1995,
McGaugh 1996) suggest that LSB galaxies are numerically common, though
considerable uncertainty remains. 

De Jong (1995, 1996) investigated the central surface brightness
distribution of spiral galaxies for a range of Hubble types.  He
concluded, on the basis of a diameter-limited, statistically complete
sample of 86 face-on spiral galaxies from the UGC (including a few
LSB galaxies), that Freeman's Law should be restated as that there is no
preferred single value for the central surface brightness of disks in
galaxies, but only an upper limit in surface brightness which galaxies
can have (corresponding approximately to the Freeman value) (see also
Allen \& Shu, 1979).  Towards
later types the central surface brightness systematically decreases. 

In general the range of properties exhibited by LSB galaxies is much
larger than that of the classical Freeman galaxies.  The sizes and
luminosities of LSB galaxies  range from the very small and
faint dwarfs like GR8 (Hodge 1967), to the giant and luminous LSB galaxy
Malin 1 (Impey \& Bothun 1989, Sprayberry et al.  1995a).  LSB galaxies
{\it in general} are therefore likely to be not so much a single group,
different from all other types of galaxies, but rather the dim
counterparts of the known ``classical'' galaxies. 

We have drawn our sample from the list of LSB galaxies by
SBSM.  The SBSM list was made by scanning by eye selected fields of the new
POSS-II survey, and applying the UGC diameter selection criteria to the
objects discovered.  The increased depth of these new plates resulted in
the discovery of many LSB galaxies that would have been in the UGC had
it been compiled at present using the POSS-II plates.  SBSM make no
claims about completeness in their survey, but rather that 
``our purpose...  is simply to increase our
known sample of LSB galaxies and to offset the clear bias in apparent
magnitude catalogs for the discovery of only HSB systems.'' The present
paper is written in this spirit. 

With this in mind it is striking that LSB galaxy surveys that have
concentrated on detecting field LSB galaxies (like e.g.  SBSM) have in general found
that the large majority of the galaxies ($\sim$ 85\% in the case of
SBSM) in these surveys are late-type spirals.  Ellipticals and
early-type spiral galaxies form only a small minority.  In our work we
will therefore restrict ourselves to describing the properties of a subset of
LSB galaxies: those of the late-type LSB disk galaxies.
In the rest of the paper when we refer to LSB galaxies we
will always implicitly mean this sub-group, unless noted otherwise.
Not included are dwarf ellipticals and spheroids, like those found in
the Local Group.

The group of late-type LSB galaxies is in many ways an extension of the
Hubble-type sequence towards very late types.  They continue (and are in
many cases the most extreme representatives known of) the trends defined
by the HSB galaxies along the Hubble sequence towards lower surface
brightnesses.  Such trends are increasingly bluer colours (BHB95,
McGaugh \& Bothun 1994 [hereafter MB94], R\"onnback 1993), decreasing
oxygen abundances in the gas (McGaugh 1994, R\"onnback \& Bergvall
1995), and decreasing H{\sc i} surface densities (van der Hulst et al.\
1993 [hereafter vdH93]).  A detailed investigation of these late-type
LSB disk galaxies will therefore provide an insight into systematic
changes of properties of spiral galaxies along the Hubble sequence as an
explicit function of surface brightness. 

One particular example of the trends with surface brightness was
presented in Zwaan et al (1995) (cf. Romanishin et al. 1982).  
It was shown that late-type LSB disk
galaxies also lie on the Tully-Fisher relation (see also Sprayberry et
al. 1995b), requiring a systematic
and fine-tuned relation between the surface brightness of a galaxy and
its overall mass-to-light ratio.  In order to hope to understand this
systematic relation, and perhaps to uncover other relations important to
our understanding of galaxy properties and their history, resolved H{\sc
i} imaging, providing detailed rotation curves, is required. 

We have obtained such H{\sc i} imaging data for 19 LSB galaxies with the
{\sc vla}\footnote{The Very Large Array of the National Radio Astronomy
Observatory (NRAO) is operated by Associated Universities Inc.  under
contract with the National Science Foundation,} and {\sc
wsrt}\footnote{The Westerbork Synthesis Radio Telescope is operated by
the Netherlands Foundation for Research in Astronomy with financial
support from the Netherlands Organization for Scientific Research
(NWO).} synthesis radio telescopes.  In this paper we present the data
and a global analysis.  In forthcoming papers we will present a full
decomposition of the rotation curves into disc and halo components, and,
in combination with optical properties, discuss the processes that
regulate the rate of evolution and star formation in these dim galaxies. 

In Section 2 we discuss the sample selection.  Section 3 contains
technical details on the observations.  In Section 4 we discuss the
reduction of the data.  In Section 5 the derivation of parameters from
these data are given.  In Section 6 we present the data in tabular and
graphical form.  Section~7 contains a discussion of the H{\sc i} surface
densities.  In Section~8 the rotation curves are discussed.  Section 9
is devoted to the mass-to-light ratios, and finally in Section 10 the
main points of this paper are summarized. 

\section{Sample Selection}

We present 21-cm H{\sc i} synthesis radio observations of
a sample of 19 LSB galaxies, taken from the catalogues of LSB galaxies
by SBSM.  
The SBSM catalogue is not dominated by small, faded dwarf galaxies. LSB
galaxies occupy the full range of types as defined by the Hubble
sequence, however with a strong preference towards later types. 

The redshift distribution is similar to those of the UGC galaxies, with
only small deviations at the low and high velocity tails, where the LSB
survey picks up LSB dwarfs, and LSB giants respectively.  The spatial
distribution shows that LSB galaxies are good tracers of the large-scale
structure on scales of $>2$ Mpc (cf. Bothun et al. 1993; Mo, McGaugh \&
Bothun 1994). 
They do not fill the voids.  The same
applies to the distribution of H{\sc i} masses: apart from a slightly
smaller mean for the LSB galaxies, they are similar.  The large fraction
of LSB galaxies that has been detected in H{\sc i} shows that these
galaxies are just as likely to contain H{\sc i} as their HSB counterparts of
similar Hubble-type.  We refer to SBSM for a complete discussion of the
properties of the galaxies in their catalogue. 

The typical LSB galaxy, as listed in SBSM, is therefore a late-type
spiral galaxy in our local universe, with luminosity and H{\sc i}
mass not significantly different from late types already present in
the UGC, but typically more extreme in surface brightness.

In order to select a number of representative galaxies from the
SBSM catalogue, we have applied the following criteria.

\begin{enumerate}
 \item The galaxy should have been detected in H{\sc i}.  This
necessarily limits the sample to the 171 galaxies that have been
observed with Arecibo, but as explained in SBSM this does not bias the
sample towards objects containing much H{\sc i}, as the success rate of
detections for these galaxies was at least 80\% (and would have been
higher if not for the limits imposed by the observable velocity range). 
 \item As the H{\sc i} masses of galaxies in SBSM are strongly peaked
around $10^9 M_{\odot}$ with a spread of 0.5 dex to both sides (cf. 
Fig.  5 in SBSM), we have selected only galaxies with H{\sc i} masses in
the range $8.5 < \log M_{\rm HI} < 9.5$.  This leaves 132 galaxies. 
 \item In order not to confuse true dwarfs with LSB disk galaxies we
have furthermore restricted the sample to those galaxies with redshifts
between 3000 and 8000 km s$^{-1}$, which SBSM define as ``LSB disks''
(cf.  their Fig.  3).  This leaves 95 galaxies. 
 \item From this sample of 95 galaxies, we selected 16 galaxies at
random. This subsample represents in our opinion a typical cross-section
of the larger sample of late-type LSB disk galaxies.  \end{enumerate}

The sample is listed in Tables 1--3.

In order to compare the properties of these LSB galaxies with those of
dwarf LSB galaxies and large LSB galaxies, we further supplemented the
sample with examples of these two classes.  The dwarf sample consists of
3 galaxies with H{\sc i} masses smaller than $10^{8.5}$ $M_{\odot}$ and
redshifts smaller than 3000 km s$^{-1}$.  These are F564-V3, F571-V2 and
F583-1. 

We used the sample of vdH93 to represent the large LSB galaxies. 
Although their H{\sc i} masses are slightly larger, they set themselves
apart from our main LSB sample mostly because of their larger optical
scale lengths, and because of the fact that despite their low central
surface brightness, they were included in the UGC catalogue.

Although our sample is not complete in a statistical sense, we have obtained
a sample of galaxies which is representative of the
late-type LSB disk galaxies found in the SBSM catalogue.  We
can use this sample to extend the range in surface brightness over which
properties of disk galaxies are investigated towards lower surface
brightnesses. 

It should also be noted that, to the best of our knowledge, no H{\sc
i}-rotation curves of galaxies with central surface brightnesses fainter
than 23.5 $B$-mag arcsec$^{-2}$ (the average surface brightness of our
sample) have been published.  There are only a handful of rotation curves
available in the literature of LSB dwarfs ($\mu_0 \sim 23$), most of
them DDO dwarfs (Carignan \& Beaulieu 1988, Hoffman et al. 1996).  This is
surprising, given the fact that these galaxies are 
dark-matter dominated, providing ideal cases for the study of the
dynamics and density distributions of the dark matter.  The discussion
on whether halos are isothermal or singular (see Flores \& Primack 1994) 
depends very much on evidence provided by this same handful of rotation
curves. 

The sample presented here thus greatly extends the available data on
galaxies in this still ill-charted LSB area of parameter space.  This is
currently the largest sample of LSB galaxies for which detailed optical,
radio and abundance data are available (vdH93, McGaugh 1994, MB94,
BHB95). 

\section{Observations}

We observed 19 LSB galaxies in the 21-cm H{\sc i} line. Observations of 10 galaxies
were performed with the {\sc vla} in its 3-km (C) configuration using
two IF systems, 27 antennas and 64 channels over a 3.125 MHz bandwidth. The
spectra were Hanning-smoothed on-line providing a velocity resolution of 
20.6 km s$^{-1}$. Each galaxy was observed for 5 times 40 minutes
interspersed with 5 minute observations of a phase calibrator. For bandpass
calibration and the flux density scale we observed 3C286 with the same
correlator settings as each galaxy. The observations were performed on 10, 16, 
17 and 21 December 1990. Further details are given in Table~1.

The other 9 galaxies were observed with the {\sc wsrt}.  Each galaxy was
observed for 12 hours with 38 baselines ranging from 72m to 2736m with
72m increments.  Both IF systems were used and the correlator was set up
to provide 64 channels over a bandwidth of 2.5 MHz.  No on-line spectral
smoothing was performed, so the highest possible velocity resolution is
9.9 km s$^{-1}$.  Off-line Hanning smoothing reduces the velocity
resolution to 16.5 km s$^{-1}$.  Amplitude/phase and bandpass
calibration was performed using observations of 3C147 and 3C286.  The
observations were done in June and September 1990 and May and August
1991. 

\begin{table*}
\begin{minipage}{130 mm}
\caption[]{List of Observations}
\begin{tabular}{llrcccc}
\hline
Galaxy  & \ \ \  R.A.  &  Dec\ \ \   &    $V_{\rm central}  $ &  Instrument &  Resolution   &   rms noise  \\
        &   (1950.0) &  (1950.0)    &    (km s$^{-1}$)   & &   (arcsec)  &    (mJy/beam)  \\
\hfil(1)\hfill&\hfil(2)\hfill&(3)\ \ \ \ &(4)&(5)&(6)&(7)\\
\hline
F563-V1  & 08 43 48.1 & 19 04 26   &  3931     &  VLA   &  13.2 $\times$ 14.1 &   1.0 \\
F563-V2  & 08 50 14.8 & 18 37 34   &  4316     &  VLA   &  13.1 $\times$ 14.0 &   1.1\\
F567-2   & 10 15 11.4 & 21 18 47   &  5683     &  VLA   &  13.4 $\times$ 14.6 &   1.2\\
F568-1   & 10 23 22.8 & 22 40 58   &  6522     &  VLA   &  13.6 $\times$ 15.5 &   1.3\\
F568-3   & 10 24 14   & 22 28 51   &  5912     &  VLA   &  12.9 $\times$ 14.0 &   1.0\\
F568-V1  & 10 42 19.9 & 22 19 20   &  5796     &  VLA   &  13.3 $\times$ 13.9 &   1.2\\
F571-V1  & 11 23 41.2 & 19 06 24   &  5705     &  VLA   &  13.4 $\times$ 14.3 &   1.1\\
F574-2   & 12 44 19   & 22 04 34   &  6322     &  VLA   &  13.6 $\times$ 15.3 &   1.6\\
F577-V1  & 13 47 56.4 & 18 31 01   &  7777     &  VLA   &  13.3 $\times$ 14.7 &   1.2\\
F579-V1  & 14 30 38   & 22 58 59   &  6302     &  VLA   &  12.9 $\times$ 14.3 &   1.1\\

F561-1   & 08 06 35   & 22 42 25   &  4810     &  WSRT  &  13.0 $\times$ 33.0 &   1.7\\
F563-1   & 08 52 10   & 19 56 35   &  3495     &  WSRT  &  12.8 $\times$ 37.1 &   1.7\\
F564-V3  & 09 00 03   & 20 16 23   & \ 480&  WSRT  &  12.6 $\times$ 36.1 &   1.5\\
F565-V2  & 09 34 47   & 21 59 46   &  3677     &  WSRT  &  12.7 $\times$ 37.2 &   1.5\\
F571-8   & 11 31 18   & 19 37 53   &  3720     &  WSRT  &  13.2 $\times$ 39.3 &   1.5\\
F571-V2  & 11 34 53   & 18 41 10   & \ 958&  WSRT  &  12.8 $\times$ 39.7 &   1.7\\
F574-1   & 12 35 38   & 22 35 20   &  6888     &  WSRT  &  13.0 $\times$ 33.0 &   1.6  \\
F583-1   & 15 55 17   & 20 48 21   &  2277     &  WSRT  &  13.0 $\times$ 35.0 &   1.6\\
F583-4   & 15 49 58   & 18 55 58   &  3621     &  WSRT  &  12.8 $\times$ 39.2 &   1.8\\
\hline
\end{tabular}

(1) Name of the galaxy\\
(2) Right Ascension of pointing center (1950.0)\\
(3) Declination of pointing center (1950.0)\\
(4) Central velocity (km s$^{-1}$)\\
(5) Instrument used to do observation\\
(6) Size of synthesized beam (arcsec)\\
(7) Root-mean-square noise (mJy beam$^{-1}$)\\

\end{minipage}
\label{Obs}
\end{table*}

\section{Reduction}
\subsection{VLA data}

The raw {\sc vla} databases of {\it uv\/}-plane visibilities were edited to
remove interference and bad baselines.  Observing conditions were generally
quite good, and very little editing was actually necessary.
The edited {\it uv\/} databases were transformed into data cubes with the
{\sc aips} routine {\sc horus}.  

Calibration was obtained from observations of standard {\sc vla} calibrators
interspersed between segments of on-target observations, typically with
several calibration observations per galaxy.  The continuum was removed
from the data cubes by subtracting the average of channels which
contained no galaxy emission.  These channels were carefully checked for
signal from the galaxies, and a buffer space of several channels was
left between the edges of the targets and the channels used to define
the continuum.  Deconvolution of sidelobes with {\sc clean} was found to
be unnecessary, as the level of the sidelobes was in general far below 
the noise level. 

The pixel size was set to $5 \times 5$ arcsec, providing good
sampling of the almost round beam of $14 \times 13$ arcsec. The
velocity resolution after Hanning smoothing was 20.6~km~s$^{-1}$.

The r.m.s.  noise in the individual, continuum-subtracted channelmaps
after Hanning smoothing was 1.1~mJy/beam, except for one galaxy (F574-2)
that was observed for a shorter period and had a r.m.s. noise in the
channel maps of 1.6~mJy/beam.  So in general the minimum detectable
$(3\sigma)$ column density in the full-resolution channel maps was $1.5
\cdot 10^{20}$ atoms cm$^{-2}$. 

\subsection{WSRT data}

The raw $uv$ data were calibrated using standard procedures (Bos et al.
1981) and Fourier-transformed to produce {\sc fits} data cubes. 
These were then reduced using the Groningen Image Processing System {\sc
gipsy} (van der Hulst et al.  1992).  The channel maps of all data
cubes were inspected visually to determine which channel maps contained
H{\sc i} line-emission.  The continuum was determined for each channel,
in the same way as for the {\sc vla} data.  A linear interpolation of
the continuum was made over the central channels and the continuum value
was subtracted.  The channel maps at the edges of the passband were not
used in making the interpolation, as these maps were unusable due to the
high noise level caused by the decreasing gain at the edges of the
passband.  This usually meant that the fits to the central continuum
values were constrained by approximately 15 to 20 channel maps. 

The resulting continuum-subtracted data cubes were Hanning-smoothed,
decreasing the effective velocity resolution  from 9.9 to 16.5~km~s$^{-1}$. 
The data cubes did not need to be {\sc clean}-ed, as the strength of the
emission was too low to show the effects of the sidelobes.  The r.m.s. 
noise in the channel maps (after Hanning smoothing) was 
1.6~mJy/beam, corresponding to a minimum detectable column density
($3\sigma$) of $1\cdot 10^{20}$ atoms cm$^{-2}$.  Note that the
better sensitivity of the {\sc wsrt} data is caused by the
longer integrations and the larger beam. 

\section{The Data}

In the rest of this paper we will refer to the number of atoms
cm$^{-2}$ (or alternatively M$_{\odot}$ pc$^{-2}$) as projected on the
sky, i.e.  {\it not corrected} for inclination, as {\it ``column
density''}, while the term {\it ``surface density''} will be used for
the number of atoms cm$^{-2}$ (M$_{\odot}$ pc$^{-2}$) {\it
corrected} for inclination effects. 

\subsection{Total column density maps and velocity fields}

In order to construct the total column density maps -- showing the
integrated H{\sc i} emission -- and the velocity fields, the data cubes
were smoothed to approximately half of their original spatial resolution
(i.e., to $40 \times 40$ arcsec for the {\sc wsrt} data and to $25
\times 25$ arcsec for the {\sc vla} data).  All pixels below 2$\sigma$
as measured in the channel maps of the smoothed cubes were blanked and
remaining spurious noise peaks were removed.  The non-blanked areas were
used as masks for the cubes at the original resolution, and only those
pixels in the original cubes that corresponded with non-blank pixels
were retained.  For each spatial position in the cube the H{\sc i}
profile moments were determined to construct
the total H{\sc i} column density maps and velocity fields. 

\subsection{Ellipse parameters}

In general the galaxies were too face-on and too small to make a good
tilted-ring model fit to the velocity fields and determine a reliable
value for the kinematical inclination.  So in most cases the inclination
was determined either from the column density map after correcting for
the shape of the beam, or the optical images (MB94, BHB95), or both.  A
check was made to see whether the value thus found was consistent with
tentative values derived from the velocity fields.  The position angles
of the major axes and the positions of the dynamical centers of the
galaxies were determined by inspecting the sequence of channel maps. 
See Table~2 for further details.  The kinematical position angle
is well determined, and usually accurate to within 5 degrees.  Errors in
the inclination are difficult to estimate, and actually depend on the
value of the inclination itself, but we are confident that the values
are accurate to at least within 5 degrees, in most cases even better. 

\subsection{Radial column density profiles and position-velocity diagrams}

Radial column density profiles were produced by integrating the column
density maps in elliptical annuli, with ellipse parameters as given in
Table~2.

The position-velocity diagrams were constructed by making a slice of
15 arcsec width through the cubes along the major axis in
the velocity direction. 

\begin{table}
\begin{minipage}{80 mm}
\caption[]{Inclination and position angle}
\begin{tabular}{lcrr}
\hline
Name    & type & $i$ & P.A.\\
(1)&(2)&(3)&(4)\\
\hline
F561-1  	&1 	&24	&55\\
F563-1  	&1,2,3	&25	&-19\\
F563-V1 	&2	&60	&-40\\
F563-V2 	&2	&29\rlap{$^a$}& -32\\ 
F564-V3		&1	&30	&156\\
F565-V2 	&1,2	&60	&-155\\
F567-2  	&2	&20	&119\\
F568-1  	&1,2,3	&26	&13\\
F568-3  	&1,2	&40	&169\\
F568-V1 	&1,2	&40	&-44\\
F571-8  	&1,2,3	&90	&165\\
F571-V1 	&1,2	&35	&45\\
F571-V2 	&2	&45	&-150\\
F574-1  	&1,2	&65	&100\\ 
F574-2  	&1,2	&30	&53\\
F577-V1 	&1,2	&35	&40\\
F579-V1 	&1,2	&26	&120\\
F583-1  	&1,2,3	&63	&-5\\
F583-4  	&1,2	&55	&115\\
\hline
\end{tabular}

(1) Name of the galaxy\\
(2) Type of inclination used for final value. `1' denotes optical
inclination, `2' H{\sc i} inclination, and `3' kinematical inclination.\\
(3) The resulting best value for the inclination in degrees.\\
(4) The position angle in degrees of the approaching side of the major axis, measured
from N through E. \\
Note:\\
 $^a$ Optical inclination is 39 degrees and well determined.  It may
   however be affected by central barlike structure, and was not used.\\
\end{minipage}
\label{Incl}
\end{table}

\subsection{Rotation curves}

The rotation curves were derived from the position-velocity diagrams by
measuring the velocities of the maxima of the H{\sc i} emission along
the major axis.  Beam-smearing was taken into account in the inner parts
of the galaxies, where the gradient in the rotation curve is largest.
The effects are however negligible (less than a few percent).
The rotation curves were each derived twice, independently by dB
and McG, and subsequently compared.  In all cases the curves agreed
within the errors.  

\begin{figure}
\epsfxsize=8.5cm  
\hfil\epsfbox{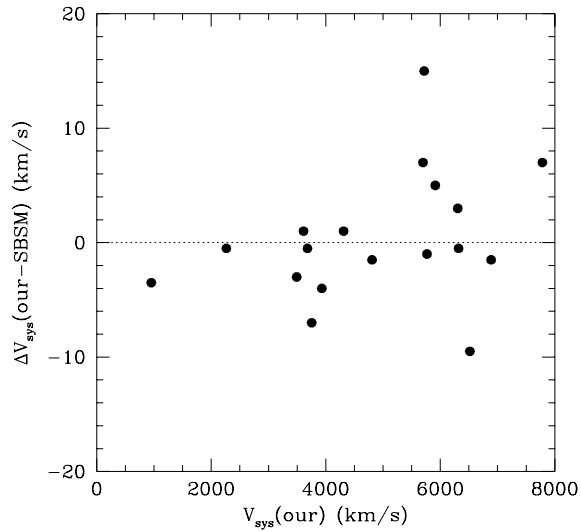}\hfil  
\caption{Difference between systemic velocities as determined from our
rotation curves and the velocities obtained by SBSM using the Arecibo
single dish telescope.}
\label{Sysvel}
\end{figure}

Fits to the rotation curves were made by using a spline fit to the
average of the symmetrized data in a plot of $|V-V_{\rm sys}|$ {\it vs}\
$|r-r_0|$.  The values of $V_{\rm sys}$ and $r_0$ were chosen to
minimize asymmetries; usually each is approximately the half way point
between the extremes of the respective variables.  The systemic
velocities determined in this way are in excellent agreement with the
single dish values from SBSM (Fig.  \ref{Sysvel}).  The mean absolute
difference is 4 km s$^{-1}$, which is less than half of one of our
channel separations. 

\section{Results}

The data are presented in Fig.~2 and in
Table~3. Following is a description of the various panels in
Fig.~2, and the columns in Table~3.

\subsection{Figures}

\subparagraph{\bf (a) R-band image with H{\sc i} overlay.}

The optical $R$-band images are shown in grayscale.  For a full
description of these images see BHB95, McGaugh et al.  (1995) and
Appendix A.  The contour levels represent H{\sc i} column densities of
$1, 3, 5, 7 ...  \times 10^{20}$ atoms cm$^{-2}$. 

The coordinate systems of the optical and radio data were aligned by
determining the right ascensions, declinations and pixel coordinates of
8 field stars in each $R$-band image.  The right ascensions and
declinations of the field stars are taken from POSS plates.  The formal
errors in aligning the two coordinate systems were at most $0.3$
arcsec. 

\subparagraph{\bf (b) B-band images}

For further details see BHB95 and McGaugh et al.  (1995). 
For galaxies F571-8, F574-1, F579-V1, F583-1 and F583-4 a $B$-band
image was not available, and a $R$-band image was used instead (see Appendix A).
Galaxy F571-V2
is not visible in the $R$-band image, due to the presence of a bright
star. We have instead retrieved an image of this galaxy from the red digital
Palomar Sky Survey which is presented here.

\subparagraph{\bf (c) H{\sc i}\ column density maps and velocity fields}

The black contours and grayscales 
represent $1, 3, 5, 7 ...  \times 10^{20}$ H{\sc i} atoms
cm$^{-2}$.  The beam shown in the lower right corner has size $13 \times
14$ arcsec for the {\sc vla} data, and on average $13 \times 35$
arcsec for the {\sc wsrt} data.  The bar in the lower left corner
shows the linear scale at the distance of the galaxies, assuming $H_0 =
100$ km s$^{-1}$ Mpc$^{-1}$. 

The velocity field is shown in white contours superimposed on the column
density map.  The velocity values
corresponding to the respective contours are given in the captions to
Fig.~2. 

\subparagraph{\bf (d) Position-velocity diagrams}

The contour levels are $-2, 2, 3, 4 ...  \sigma$, where $\sigma$ is the
standard deviation in the velocity slice.  One $\sigma$ equals 1.7
mJy/beam for the {\sc vla} data and 2.5 mJy/beam for the {\sc wsrt} data. 
The horizontal dashed line shows the systemic velocity as derived from
our symmetrization of the rotation curve, while the vertical line shows
the spatial position of the center of rotation, also derived from
symmetrization.

\subparagraph{\bf (e) Radial surface density profiles}

Radial column density profiles were determined from averaging in
elliptical annuli (``rings'' in the plane of the galaxy).  These were
corrected for inclination.  Shown are therefore the radial {\it surface
density} profiles.  The error bars indicate the uncertainties in the
values, and their size is given by:

\[ \sigma_{\rm tot} = \sigma_{\rm ch}\cdot {{4} \over{\sqrt{6}}}\cdot
{{\sqrt{N-0.75}}\over {\sqrt{\rm beams/ring}}} \]
where $N$ is the number of channels contributing to a pixel in the
total column density map ($N\simeq 3$), and $\sigma_{\rm ch}$ is the r.m.s.
noise in a single channel map. 

The lower horizontal axis gives the linear distance in kpc, using the
distances from Table 3; the upper horizontal axis shows the
corresponding angular distance in arcseconds.

In the case of F571-8 we show both the {\it observed} (edge-on) column
density distribution and the deduced face-on surface density
distribution.  As F571-8 is an edge-on galaxy, deriving the latter
distribution is not straight-forward.  We have made an attempt, shown in
the inset in panel (e) of the F571-8 panels, by dividing the galaxy into
concentric rings (as seen face-on), assuming that gas is everywhere
present at the line of nodes, and by working our way from the outside
in, subtracting the contributions of subsequent rings, taking into
account the lines of sight through each particular ring.  This assumes
of course that the H{\sc i} is optically thin everywhere. 

\subparagraph{\bf (f) Rotation curves}

The rotation curve data shown here are corrected for inclination.
The drawn line is the symmetrical approximation to the
data.  See Section 5.4 for more detailed information. The open dots
always correspond to the receding side of the galaxy, while the filled dots
correspond to the approaching side. Again the bottom axis shows the
radius in kpc, while the top axis shows the angular distance in
arcseconds. The arrow indicates one optical disk scale length.

\subsection{Table}

Following is a description of the contents of Table 3.

{\bf Column (1):} The name of the galaxy (SBSM).

{\bf Column (2):} Heliocentric systemic velocity $V_{\rm sys}$, defined
as the center
velocity in km s$^{-1}$ of the symmetrized rotation curves. 

{\bf Column (3):} Distance $D$ in Mpc, as derived from the
heliocentric systemic velocity, given in Column (2), after
correction for Galactic rotation and a Virgocentric flow of 300~km~s$^{-1}$
(where necessary) and assuming $H_0 = 100$ km s$^{-1}$ Mpc$^{-1}$.

{\bf Column (4):} Logarithm of the total H{\sc i} mass $M_{\rm HI}$.  The total
H{\sc i}-mass, given in $M_{\sun}$, is derived from the total flux using the
formula
 \[ M_{\rm HI}/M_{\odot} = 2.36 \cdot 10^{5} D({\rm Mpc})^2 \int S({\rm
Jy})dV({\rm km\ s}^{-1})  \] 
 The fluxes are in good agreement with the 
single dish fluxes from SBSM.  Figure \ref{Fluxes} compares the synthesis
and single dish fluxes, and shows that the synthesis observations measure
all the flux.

{\bf Column (5):} Maximum rotation velocity $V_{\rm max}$ of the rotation
curve, not corrected for inclination. 

{\bf Column (6):} Outer rotation velocity $V_{\rm out}$.  This is the
rotation velocity in km s$^{-1}$ as measured at the outermost point of
the rotation curve.  This value is not corrected for inclination. 

{\bf Column (7):} Outermost radius $R_{\rm out}$.  This is the radius in
arcsec where $V_{\rm out}$ is measured and is the maximum radius at
which the rotation curve could be measured. 

{\bf Column (8):} Logarithm of the dynamical mass $M_{\rm dyn}$
encompassed at the last measured point.  This
dynamical mass was derived using the formula
\[ M_{\rm dyn} = {{V_{\rm out}^2 R_{\rm out}}\over {G \sin^2 i}}, \]
where $R_{\rm out}$ is expressed in linear units.

{\bf Column (9):} Ratio $M_{\rm HI}/M_{\rm dyn}$, the ratio of H{\sc i} mass and
dynamical mass. 

{\bf Column (10):} H{\sc i} radius $R_{\rm HI}$.  The radius $R_{\rm HI}$ is defined
as the radius where the azimuthally averaged {\it surface} density falls below 1
$M_{\sun}$ pc$^{-2}$.  Note that this radius is measured in the
inclination corrected radial {\it surface} density profile. As F571-8 is
an edge-on galaxy, the value given here for that galaxy should not be directly compared
with the values for the other galaxies.

{\bf Column (11):} Radius in arcseconds of the 25 $B$-mag arcsec$^{-2}$
isophote $R_{25}$.

{\bf Column (12):} Adopted inclination (see Table 2).

{\bf Column (13):} Peak smoothed column density $\sigma_{\rm smo}$.  In
order to better compare the peak surface densities as measured with the
{\sc vla} and the {\sc wsrt}, the {\sc vla} data was smoothed to $35
\times 13$ arcseconds resolution (which is the {\sc wsrt} resolution) 
and the resulting peak column densities are given. 

{\bf Column (14):} Peak column density $\sigma_{\rm peak}$.  This is the
peak column density in each galaxy expressed in $M_{\sun}$ pc$^{-2}$. 

\setcounter{figure}{2}
\begin{table*}
\begin{minipage}{150mm}
\caption[]{Properties of the sample of LSB galaxies.}
\begin{tabular}{lrrrrrrrrrrr}
\hline
Name    &$V_{sys}$&$D$&$M_{\rm HI}$&$V_{max}$&$V_{out}$&$R_{out}$&$M_{dyn}$&${M_{\rm HI}\over M_{dyn}}$&$R_{\rm HI}$&$R_{25}$&$i$\\
(1)&(2)&(3)&(4)&(5)&(6)&(7)&(8)&(9)&(10)&(11)&(12)\\
\hline
 F561-1 &  4809& 47& 8.91& 21& 21& 33& 9.66& 0.177& 33.2&21&24\\
 F563-1 &  3492& 34& 9.19& 47& 47& 81&10.58& 0.041& 99.0&23&25\\
 F563-V1&  3932& 38& 8.48& 26& 24& 30& 9.00& 0.283& 26.1&13&60\\
 F563-V2&  4312& 46& 9.11& 54& 54& 31&10.30& 0.064& 39.3&23&29\\
 F564-V3&   476&  6& 7.11& * &  *&  *&    *&     *& 37.8&11&30\\
 F565-V2&  3681& 36& 8.53& 44& 44& 36& 9.58& 0.089& 29.1&10&60\\
 F567-2 &  5698& 56& 9.09& 22& 22& 31& 9.91& 0.152& 39.0&9&20\\
 F568-1 &  6517& 64& 9.35& 52& 51& 36&10.55& 0.064& 37.1&18&26\\
 F568-3 &  5913& 58& 9.20& 77& 77& 44&10.62& 0.038& 40.4&21&40\\
 F568-V1&  5768& 60& 9.14& 80& 80& 49&10.71& 0.027& 36.8&16&40\\
 F571-8 &  3754& 36& 8.91&133&133& 67&10.68& 0.017& (95)& 33&90\\
 F571-V1&  5719& 59& 8.82& 42& 42& 38&10.13& 0.049& 26.0&10&35\\
 F571-V2&   955& 12& 7.81& 32& 32& 47& 9.12& 0.049& 41.1&*&45\\
 F574-1 &  6889& 72& 9.29& 91& 91& 33&10.43& 0.072& 34.5&38&65\\
 F574-2 &  6319& 66& 8.97& 20& 20& 25& 9.50& 0.293& 29.2&9&30\\
 F577-V1&  7783& 80& 9.15& 17& 16& 23& 9.20& 0.891& 28.4&17&35\\
 F579-V1&  6305& 64& 9.06& 52& 44& 42&10.49& 0.037& 48.3&27&26\\
 F583-1 &  2264& 24& 8.99& 79& 79& 94&10.27& 0.052& 91.2&21&63\\
 F583-4 &  3612& 37& 8.49& 55& 55& 42&9.90& 0.0389& 30.2&27&55\\
\hline
Name&$\sigma_{smo}$&$\sigma_{peak}$&$\sigma_{aver}$&$\mu_B(0)$&$\alpha_B$&$R_{\rm HI}\over R_{25}$&$M_B$&$L_B$&$M_{\rm HI}\over L_B$&$M_{dyn}\over L_B$&slope\\
 &(13)&(14)&(15)&(16)&(17)&(18)&(19)&(20)&(21)&(22)&(23)\\
\hline
 F561-1 &5.88&5.88 & 4.80& 23.28&11.9& 1.58&--17.2& 9.08&0.68& 3.9&0.144\\
 F563-1 &6.28&6.28 & 4.94& 23.63&19.7& 4.30&--16.7& 8.88&2.06&50.1&0.186\\
 F563-V1&6.45&8.91 & 6.88& 24.30& 9.7& 2.01&--15.7& 8.49&0.91& 3.2&0.209\\
 F563-V2&8.89&14.46& 7.92& 22.10& 7.0& 1.71&--17.6& 9.23&0.76&11.8&0.204\\
 F564-V3&4.82&4.82 & 2.56& 24.05&11.2& 3.60&--12.1& 7.04&1.18&    *&   *\\
 F565-V2&7.13&7.13 & 6.16& 24.73&11.5& 2.91&--14.8& 8.11&2.62&29.4&0.607\\
 F567-2 &4.83&7.6  & 4.23& 24.42&15.8& 4.33&--16.8& 8.90&1.56&10.3&0.251\\
 F568-1 &8.63&13.39& 8.93& 23.77&12.9& 2.06&--17.5& 9.19&1.45&22.8&0.093\\
 F568-3 &6.79&8.67 & 6.09& 23.08&10.8& 1.92&--17.7& 9.27&0.86&22.5&0.324\\
 F568-V1&7.04&10.32& 5.24& 23.30& 8.2& 2.30&--17.3& 9.10&1.11&41.1&0.293\\
 F571-8 &21.35&21.35&20.41&22.12&16.0& 2.87&--17.0& 8.99& 0.83&49.0&0.575\\
 F571-V1&4.85&7.21 & 6.05& 24.00& 8.4& 2.61&--16.4& 8.76&1.14&23.4&0.345\\
 F571-V2&8.09&8.09 & 5.05&     *&   *&    *&     *&    *&    *&    *&0.429\\
 F574-1 &10.33&10.33& 6.81&23.31&10.0& 0.91&--17.8& 9.32&0.93&12.9&0.714\\
 F574-2 &4.32&7.84 & 4.36& 24.35&14.0& 3.24&--17.0& 9.00&0.93& 3.2&0.665\\
 F577-V1&5.83&8.81 & 5.00& 23.95&11.1& 1.67&--17.6& 9.23&0.83& 0.9&-0.085\\
 F579-V1&4.53&7.65 & 3.99& 22.83&10.0& 1.79&--18.2& 9.48&0.38&10.2&-0.015\\
 F583-1 &14.18&14.18&10.12&24.03&10.2& 4.34&--15.9& 8.55&2.74&52.3&0.368\\
 F583-4 &3.99 &3.99 & 3.14&23.76&11.3& 1.11 &--16.3& 8.72&0.59&15.1&0.747\\
\hline
\end{tabular}
\end{minipage}
\end{table*}

\begin{table*}
\begin{minipage}{150mm}
\begin{tabular}{ll}
(1) Name of galaxy&(13) Smoothed peak column density ($M_{\sun}$~pc$^{-2}$)\\
(2) Systemic velocity (km s$^{-1}$)&(14) Actual peak column density ($M_{\sun}$~pc$^{-2}$)\\
(3) Distance (Mpc)& (15) Radial peak column density ($M_{\sun}$~pc$^{-2}$)\\
(4) Logarithm of H{\sc i} mass ($M_{\sun}$)& (16) Central surface
brightness disc ($B$-mag arcsec$^{-2}$)\\
(5) Max. uncorrected rotation velocity (km s$^{-1}$)&
(17) Optical scale length in $B$ (arcseconds)\\
(6) Outer uncorrected rotation velocity (km s$^{-1}$)&(18)
Ratio H{\sc i} diameter and optical diameter\\
(7) Radius outermost point rotation curve (arcsec)&(19)
Absolute magnitude ($B$-mag)\\
(8) Logarithm of dynamical mass ($M_{\sun}$)&(20)
Logarithm of luminosity in $B$ ($L_{\sun}^B$)\\
(9) Ratio H{\sc i} mass to dynamical mass&(21) H{\sc i} mass to light ratio ($M_{\sun}/L_{\sun}^B$)\\
(10) Radius 1 $M_{\sun}$~pc$^{-2}$ H{\sc i} level (arcsec)&(22) Dynamical mass
to light ratio ($M_{\sun}/L_{\sun}^B$) \\
(11) Radius 25 $B$-mag arcsec$^{-2}$ isophote (arcsec)&(23) Slope of
rotation curve\\
(12) Inclination (see Table 2)&\\
\end{tabular}
\end{minipage}
\end{table*}

{\bf Column (15):} Peak radial averaged surface density $\sigma_{\rm aver}$. 
This is the maximum of the radial profile in $M_{\sun}$~pc$^{-2}$. 

The following columns contain a number of parameters related to the
optical properties of the galaxies.  These are taken from MB94 or BHB95.
The optical properties of F571-8, F574-1, F579-V1 and F583-4 are
converted from $R$-band measurements.  See Appendix A for a description. 

{\bf Column (16):} Central $B$-band surface brightness $\mu_0$ of the
exponential disc, expressed in mag arcsec$^{-2}$, corrected for
inclination, but not for internal extinction. 

{\bf Column (17):} Optical exponential $B$-band scale length $\alpha$ in
arcseconds.

{\bf Column (18):} Ratio of the optical and the H{\sc i} diameter
$R_{\rm HI}/R_{25}$. 

{\bf Column (19):} Absolute magnitude $M_{\rm abs}$, using $D$ as given in
column (3).

{\bf Column (20):} Total intrinsic luminosity $L_B$, expressed in
$L_{\sun}$, derived from $M_{\rm abs}$, assuming $M_{\sun ,B} = 5.48$.

{\bf Column (21):} Total hydrogen-mass to total light ratio
$M_{\rm HI}/L_B$, expressed in solar units.

{\bf Column (22):} Total dynamical-mass to light ratio $M_{\rm dyn}/L_B$ in
solar units.
\begin{figure}
\epsfxsize=8.5cm
\hfil\epsfbox{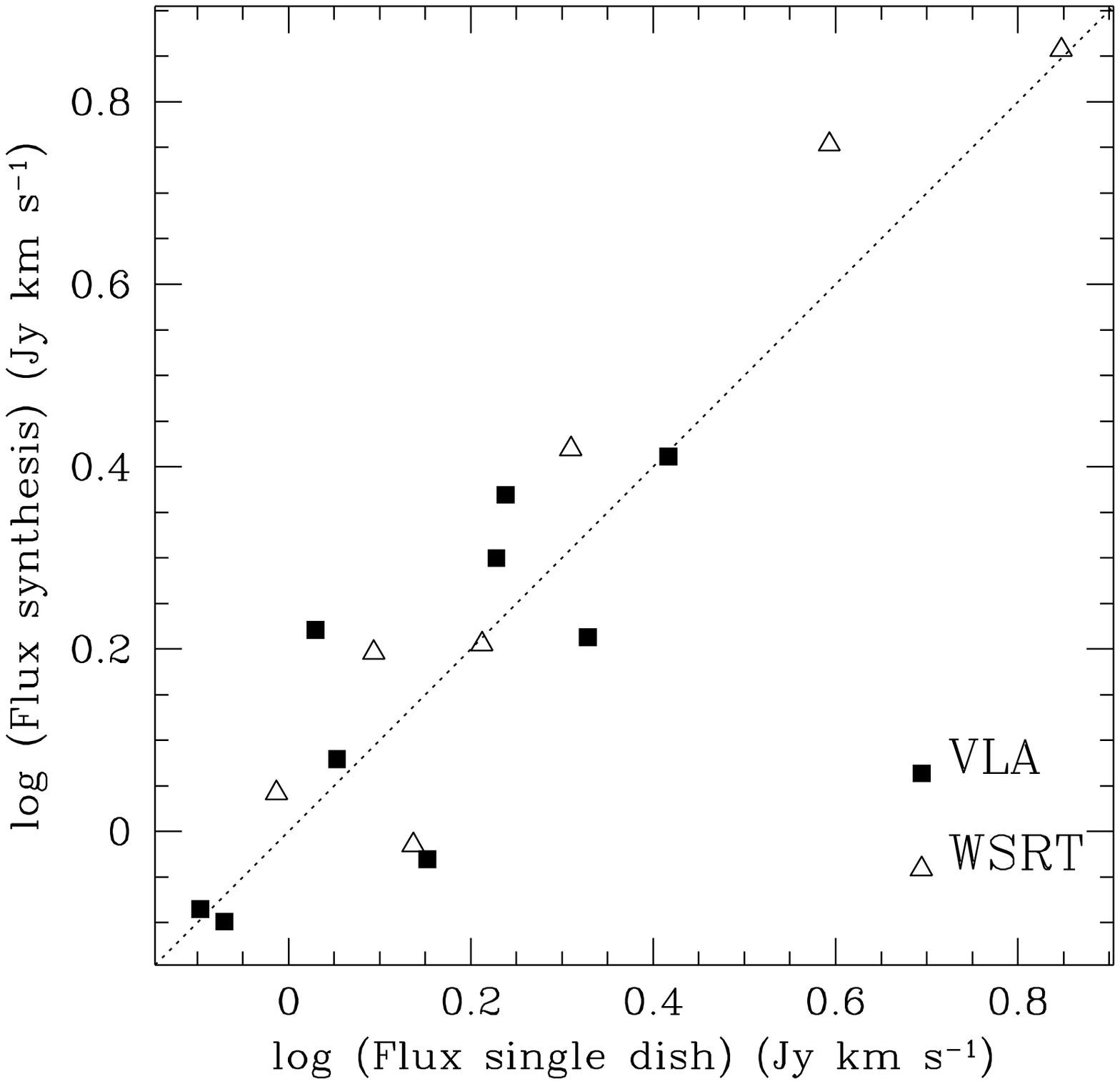}\hfil
\caption{Total flux of the H{\sc i} emission as measured with the VLA and WSRT plotted
against the flux measured with the Arecibo single dish telescope (SBSM).
The dotted line is the line of equality.}
\label{Fluxes}
\end{figure}

{\bf Column (23):} The outer slope of the rotation curve.  See
Section~8.1 for more details. 
\setcounter{page}{27}

\section{Surface densities}
In this section we will compare our data with samples of HSB galaxies,
and show that indeed LSB galaxies have systematically lower H{\sc i}
surface densities than HSB galaxies.

It is a well-known fact that properties of galaxies change along the
Hubble sequence.  One of these properties is the average surface density
of the atomic hydrogen $\sigma_{\rm HI}$.  This changes from a value of
$\sim 1$ M$_{\sun}~{\rm pc}^{-2}$ for Sa galaxies to a value of $\sim 8$
M$_{\sun}~{\rm pc}^{-2}$ for late-type galaxies (Roberts \& Haynes 1994). 
Observations by vdH93 did show that LSB galaxies do not follow this
trend.  The $\sigma_{\rm HI}$ of the LSB galaxies investigated in vdH93
is a factor of two lower than in HSB galaxies of comparable
morphological type. Note that this does not mean that LSB galaxies are
therefore also more gas-poor. On the contrary, the gas fraction is
significantly higher than in HSB galaxies, as we will show in Section~9.

Here we will compare the average and local H{\sc i} properties of HSB
and LSB galaxies, and discuss the relation between surface brightness
and surface density -- the relation between stars formed, and material
from which to form stars -- which can tell us something about the
evolution of LSB galaxies. 

\subsection{Average surface densities}

In order to make a comparison between average H{\sc i} surface densities
of HSB and LSB galaxies we need a sample of undisturbed HSB galaxies. 
Cayatte et al.\ (1994) have composed such a sample.  Their field galaxy
comparison sample consists of 84 galaxies with data taken from Warmels
(1988) and Broeils (1992).  Galaxies with large scale disturbances
either in radio or optical were excluded and only galaxies with Hubble
type between S0 and Sdm and inclinations smaller than 80 degrees were
chosen.  This sample therefore constitutes a sample of undisturbed HSB
normal Hubble sequence galaxies, with central surface brightnesses
typically between 21 and 23 $B$-mag arcsec$^{-2}$, i.e.  from the
Freeman value down to values approaching those of the LSB sample.  One
should therefore not think of the HSB sample as a collection of Freeman
galaxies.  The typical surface brightness of a typical galaxy from the
HSB sample is, however, some two magnitudes brighter than that of a
typical galaxy from the LSB sample.  
The distinction between HSB and LSB galaxies is merely a convenient way
to describe two subsets of the continuum of galaxies. 

\begin{figure}
\epsfxsize=8.5cm
\hfil\epsfbox{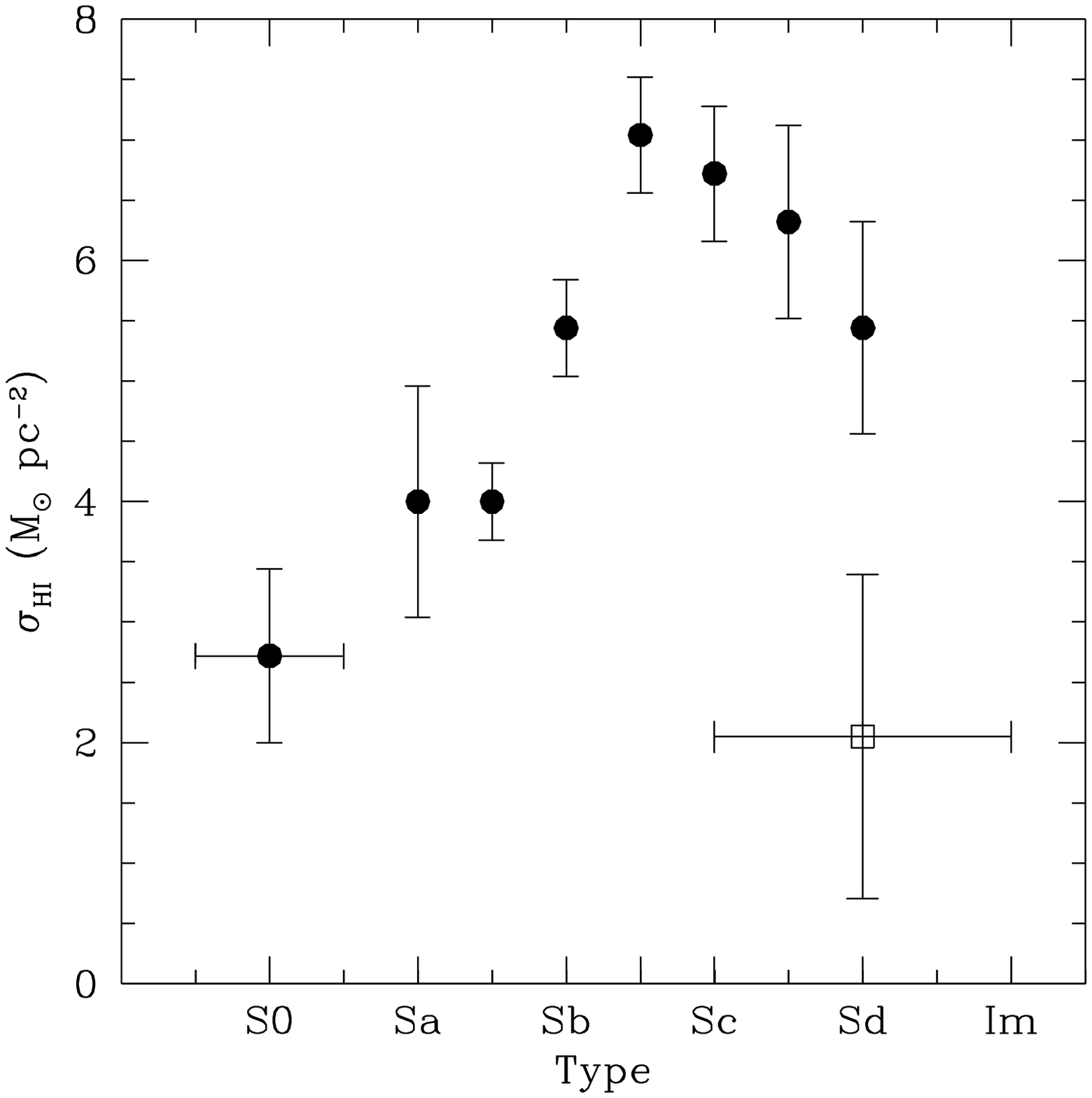}\hfil
\caption{LSB galaxies and their place in the Hubble-type -- average
H{\sc i} surface
density diagram. The average surface density is defined as the average
surface density of the H{\sc i} within half of the optical radius
$R_{25}$. Filled circles are HSB galaxies from the sample of
Cayatte et al. (1994). Vertical errorbars are the 1$\sigma$ errors in
the mean. The open square represents our sample of LSB galaxies.
Horizontal errorbars denote the range in type in our sample, vertical
errorbars are $1\sigma$ deviations from the mean. }
\label{surf}
\end{figure}

Cayatte et al.\ defined the average H{\sc i} surface density
$\sigma_{\rm HI}$ as the
average surface density of the H{\sc i} within half of the optical
radius $R_{25}$.  The distribution of average surface density with
Hubble-type is shown in Fig.  \ref{surf}.  The surface density increases
from Sa to Sc, and then slightly declines towards Sm galaxies. 

We have computed $\sigma_{\rm HI}$ for our LSB galaxies in a similar
manner.  The average surface density of LSB galaxies is also shown in
Fig.  \ref{surf}.  It is clear that late-type LSB galaxies in our sample
have average H{\sc i} surface densities which are on average a factor of
3 lower than those of HSB galaxies of comparable type. 

If we regard LSB galaxies as the continuation of the Hubble sequence
towards extremely late types, LSB galaxies would simply continue
the trend of decreasing H{\sc i} surface density seen in Fig. 
\ref{surf} from type Sc onwards. 

The Hubble sequence is then a sequence of increasing atomic gas surface
density from S0 to Sc, and decreasing surface density for the later
types.  A very interesting question is whether this decreasing trend
continues towards progressively lower surface brightnesses.  At a
certain point the gas will be diluted enough for the intergalactic
radiation field to completely ionize it.  Calculations by e.g. 
Dalcanton, Spergel \& Summers (1995) and Maloney (1993) show that this
can already happen at H{\sc i}-surface densities of a few times
10$^{19}$ cm$^{-2}$.  This would imply a minimum
surface brightness a galaxy can have below which it will become completely
ionized (Quinn, Katz \& Efstathiou 1995).

One could furthermore speculate that the low gas surface densities
measured for types earlier than Sc are caused by the rapid gas
consumption due to the high star formation rate in these galaxies (see
e.g.  Guiderdoni \& Rocca-Volmerange 1987).  In this picture the peak in
Fig.  \ref{surf} would then shift towards later types as time (and gas
consumption, and other evolutionary processes) progresses. 

\subsubsection{Isophotal ratios}

As was already remarked in MB94 and BHB95, one should be  cautious
in using isophotal quantities for a direct comparison between HSB and
LSB galaxies.  There is therefore also a danger in using optical radii to
determine average quantities, like the average surface density within
half an optical radius, as was done above.  Quantities like $R_{25}$
become less representative as one goes towards lower surface
brightnesses, and may loose their meaning altogether (in this specific
case, if the galaxy has a central surface brightness fainter than 25 mag
arcsec$^{-2}$). 

This of course also applies to ratios of such quantities, like the ratio
$R_{\rm HI}/R_{25}$ (see column~(18) of Table~3).  This is not a very
useful quantity for a quantitative comparison of HSB and LSB galaxies. 
The low surface brightness of LSB galaxies ensures that the $R_{25}$
will substantially underestimate the real optical size of the galaxy,
and artificially raises the value of the ratio.  It is obvious that
quantities like $\sigma_{\rm HI}$, as defined above, will then also be
affected. 

In this case $R_{25}$ will, however, always decrease faster with
decreasing central surface brightness than any of the H{\sc i}
parameters.  This alone will already artificially raise the computed
average surface density, so that the discrepancy between HSB and LSB
galaxies might in reality be larger than suggested by Fig.  \ref{surf}. 
We will come back to this in the next section. 

$R_{\rm HI}$ can also suffer from the same problem as $R_{25}$: as the
H{\sc i} surface density in LSB galaxies is much lower than in HSB
galaxies, it will underestimate the size of the H{\sc i} disc.  This is
apparent if one compares the distribution of values of $R_{\rm HI}$ of
HSB galaxies with that of LSB galaxies.  These distributions are
comparable, which one naively would not expect as LSB galaxies are more
extended.  However, as the surface density is lower in LSB galaxies,
this simply means that, for a HSB and a LSB galaxy with the same total
H{\sc i}-mass and $R_{\rm HI}$, there is more H{\sc i} present outside
$R_{\rm HI}$ in the LSB galaxy.  So comparing galaxies with identical
H{\sc i} mass {\it inside} $R_{\rm HI}$ is not correct, as one then
compares a more massive LSB galaxy with a lower mass HSB galaxy. 

This once again shows that isophote and iso-density effects should
always be taken into account when comparing galaxies of different
surface brightnesses and surface densities. 

\subsection{Radial density profiles}

Ideally, we would like to see if LSB galaxies have consistently
lower surface densities at all radii.  For that we need to compare the
radial H{\sc i} profiles of LSB galaxies with those of a sample of
``classical'' field galaxies. 

For this we again use the sample of Cayatte et al. (1994) as they
have calculated the average H{\sc i} surface density profiles of
galaxies of different Hubble-types.

\begin{figure}
\epsfxsize=8.5cm
\hfil\epsfbox{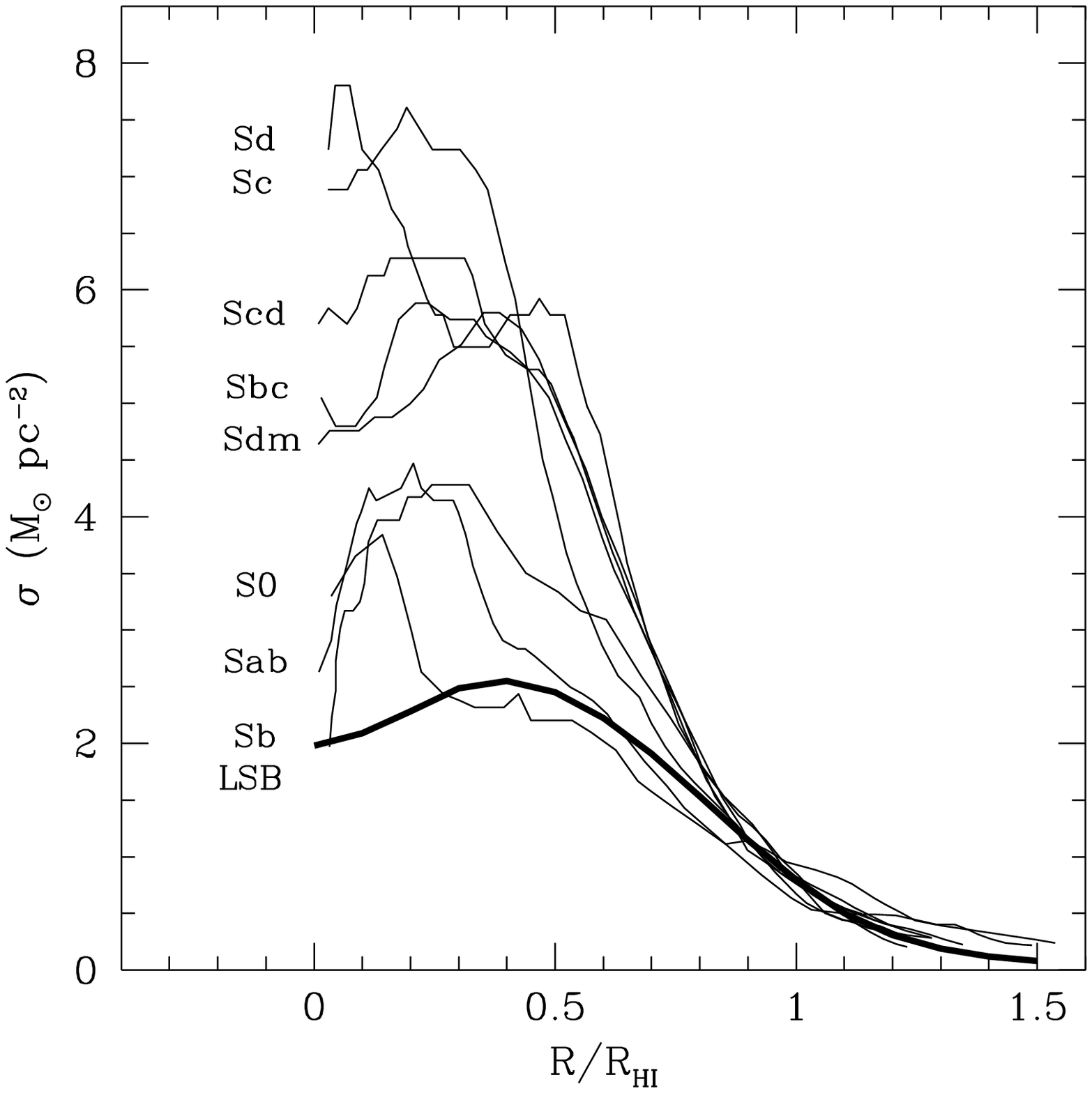}\hfil
\caption{The average radial distribution of H{\sc i} for HSB galaxies of
different Hubble-types (light
lines) (Cayette et al.\  1994) and LSB galaxies (heavy line). Radii have
been normalized to $R_{\rm HI}$ (see text).}
\label{Radprof}
\end{figure}

Unfortunately they give the average radial H{\sc i} profile for each
Hubble-type as a function of $R/R_{25}$.  As isophotal effects are more severe
in the optical than in H{\sc i}, we have re-normalized the radial
profiles of the HSB galaxies to $R/R_{\rm HI}$, using the mean values of
$R_{25}/R_{\rm HI}$ as given in Cayatte et al.\ (1994). We have made a
correction for the different definition they use for $R_{\rm HI}$. 

\begin{figure}
\epsfxsize=8.5cm
\hfil\epsfbox{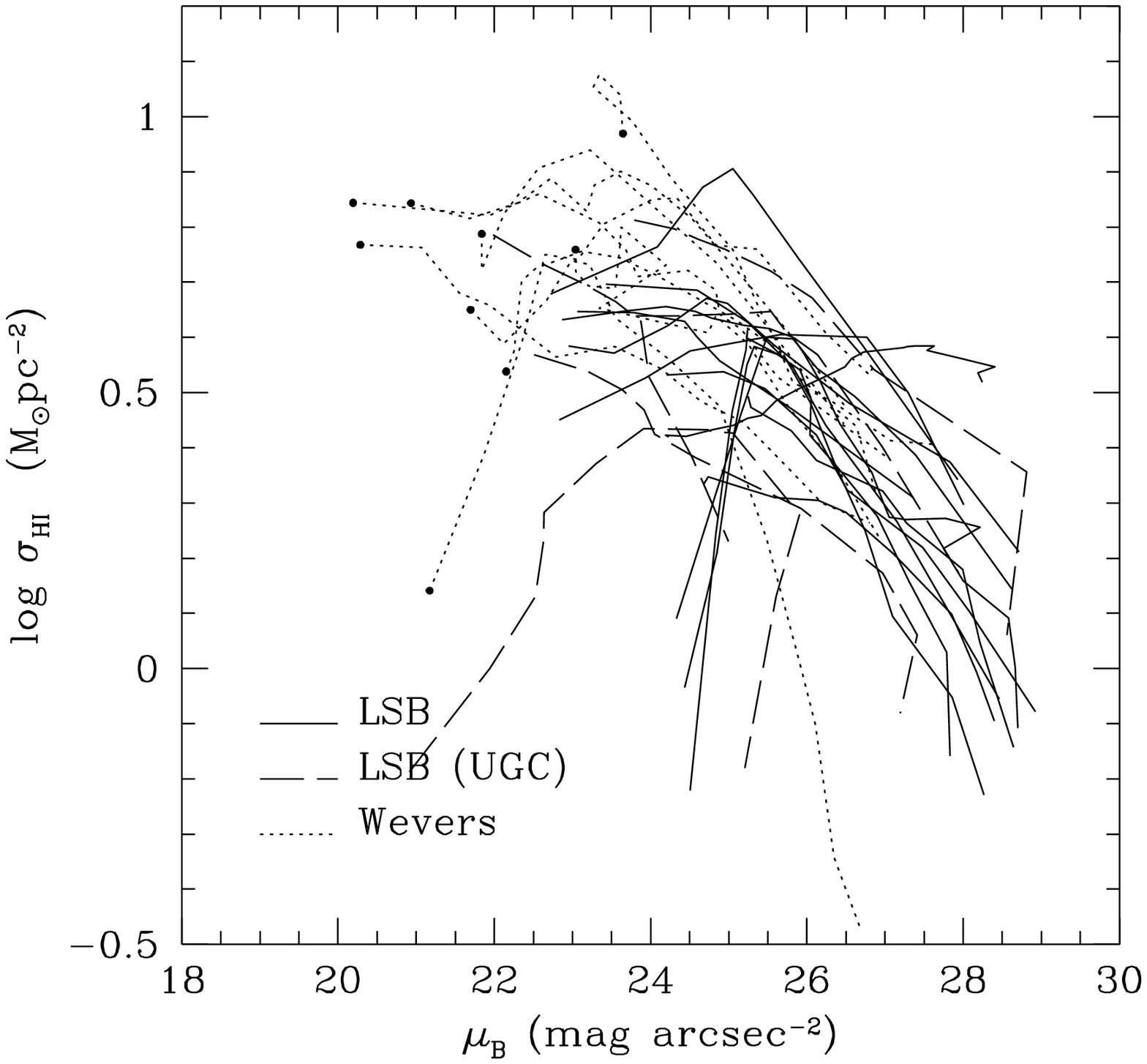}\hfil
\caption{The relation between local optical surface brightness and local H{\sc i}
surface density. The HSB galaxies from Wevers et al. (1986) are represented by
dotted lines, LSB galaxies from our sample by the heavy, drawn lines,
and the LSB galaxies from vdH93 by dashed lines.}
\label{Wevers}
\end{figure}

These profiles are presented in Fig.  \ref{Radprof}.  Note that the
trend shown in Fig.  \ref{surf} of surface density with Hubble-type is
also apparent in the local surface densities.  Local surface density
values increase from early types to Sc/Sd and then decrease again.  The
average radial surface density profile of our sample of LSB galaxies is
shown as the heavy line in Fig.  \ref{Radprof}. 

It is clear from Fig.  \ref{Radprof} that the H{\sc i} surface density
of an average late-type LSB galaxy is lower than or comparable to the
profiles of early-type HSB galaxies, and again a factor of 3 lower than
those of late-type galaxies.  This of course requires a LSB galaxies of
a given H{\sc i}-mass to be more extended than HSB galaxies of the same
H{\sc i} mass. 

\subsection{Surface brightness and surface density}

\begin{figure}
\epsfxsize=8.5cm
\hfil\epsfbox{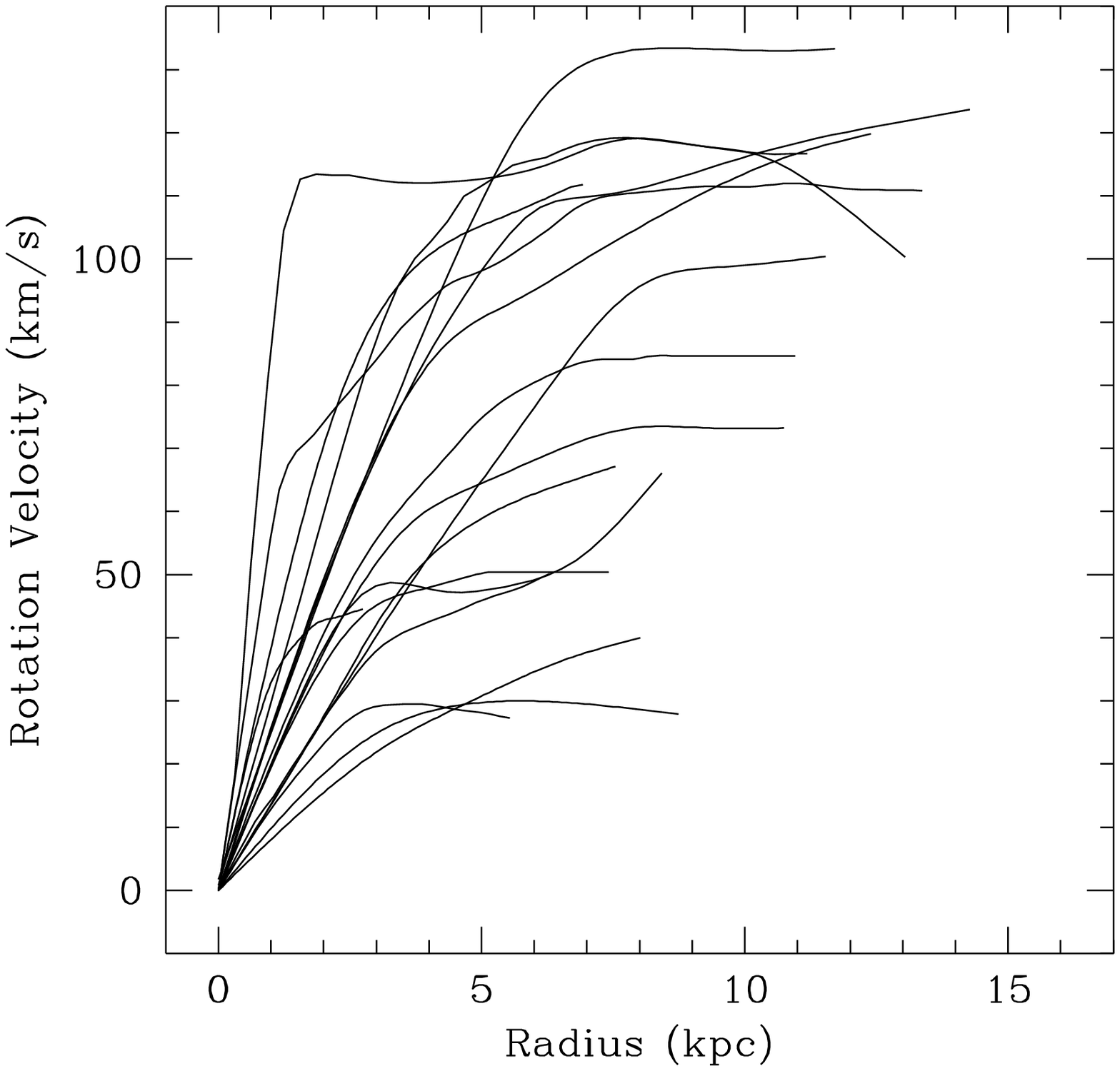}\hfil
\caption{Inclination corrected rotation curves of our sample of LSB galaxies.}
\label{Rotkpc}
\end{figure}

The result that LSB galaxies, apart from having lower surface
brightnesses, also have lower H{\sc i} surface densities shows that
there is a relation between the luminosity surface density and the gas
surface density.  We have tested this idea by taking the photometry and
the H{\sc i} data from the sample of Wevers et al.  (1986), where we
have converted his $J$-band photometry to $B$-band photometry using the
conversion formulae given by him.  The maximum error in the conversion
formulae is $\pm 0.10$ mag.  However, the uncertainties in the surface
brightness are dominated by the uncertainties in the flux-calibration,
so the surface brightnesses of the Wevers galaxies are uncertain by $\pm
0.20$ mag arcsec$^{-2}$.  As we will see this is of sufficient quality
to make a comparison with our photometry. 

Figure \ref{Wevers} shows the result.  Here local H{\sc i} surface
density at a particular radius is plotted against local surface
brightness at that radius .  Early-type galaxies (Sa--Sbc) from the
Wevers et al.  sample have been omitted, in order to avoid comparing
very different types of galaxies. 

Many of the galaxies show a central depression, but it is clear that
outside these central dips the surface density-surface brightness
relations for the different galaxies looks similar.  

Note that U6614 (the dashed line in the lower-left of the figure) turns
out to be the odd galaxy in the LSB samples.  It has high surface
brightnesses in the inner parts and low H{\sc i} surface densities,
showing that these Malin-1-cousins belong to a different class of LSB
galaxies.  The early-type galaxies from the Wevers sample, that have
been omitted from this plot, also show similar profiles. 

Figure \ref{Wevers} shows that the normal LSB galaxies follow a
relationship between $\sigma_{\rm HI}$ and $\mu_B$ that is similar to that of
late-type HSB galaxies (albeit starting at a lower $\sigma_{\rm HI}$), suggesting
that LSB galaxies as a whole may be similar to the outer parts of
late-type HSB galaxies, and that {\it local} baryonic surface density
controls the current evolutionary rate. 

However, one should keep in mind that at identical {\it local} surface
brightnesses LSB galaxies are in general bluer (de Jong 1995) and have
lower abundances (McGaugh 1994) than HSB galaxies.  This is most likely
an effect of the different evolutionary history of LSB galaxies. 
Summarizing, {\it local} surface density controls the current
evolutionary rate, i.e.  the rate at which processes happen at this
moment, while the {\it global} properties show us the total amount of
evolution that a galaxy has undergone over its lifetime.  The change in
global properties is therefore ultimately governed by {\it local}
processes. 

\section{Rotation curves}

In order to intercompare the rotation curves in Fig.  2 we present them
again in Fig.  \ref{Rotkpc}.  The rotation velocity is given as a
function of linear distance from the centre.  Velocities and linear
distances along the major axis have been computed using the inclinations
and distances given in Tables 2 and 3.  It is clear that most of the
rotation curves have maximum rotation velocities between 50 and 120 km
s$^{-1}$.  Some of the LSB galaxies are among the slowest rotators known
($\sim 30$~km~s$^{-1}$).  Only a few of the rotation curves show clear
signs of flattening off.  Usually the velocity is still increasing at
the outermost measured point. 

In order to compare the rotation curves of LSB galaxies with those of
other galaxies we have retrieved a number of well-determined rotation
curves of HSB galaxies from the literature (see Broeils 1992 and
references therein and Table A3). 

\begin{figure*}
\epsfxsize=17cm
\hfil\epsfbox{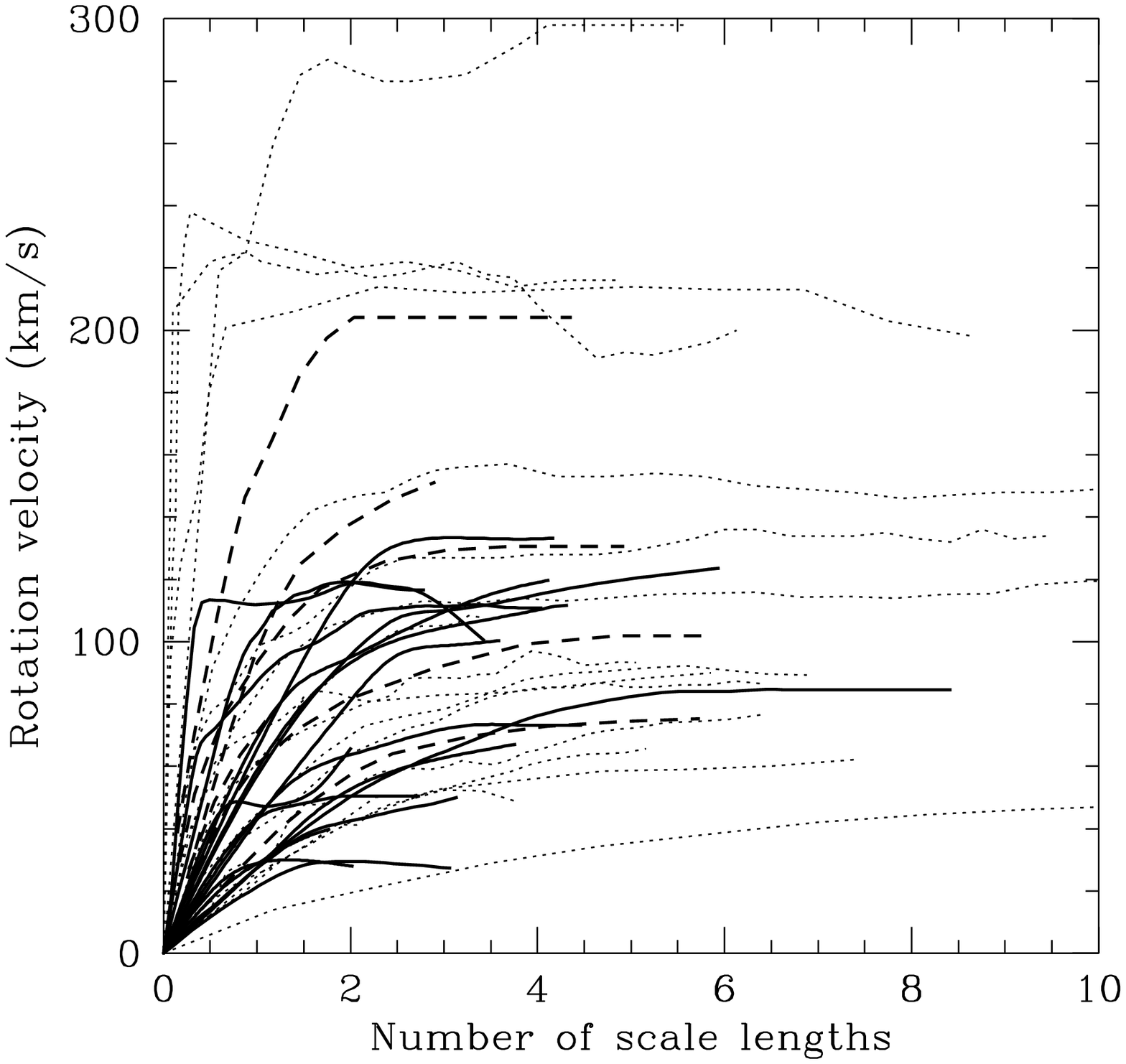}\hfil
\caption{Rotation curves of LSB galaxies from our sample (full lines),
LSB galaxies from vdH93 (dashed lines) and 
HSB galaxies from Broeils (1992) (dotted
lines), plotted versus the number
of scale lengths.}
\label{Rot_HSB_LSB}
\end{figure*}

The rotation curves are shown in Fig.  \ref{Rot_HSB_LSB}, together with
the rotation curves from Broeils.  In order to compare only galaxies
that are morphologically similar to LSB galaxies we have plotted only
HSB galaxies with Hubble-types Sc or later.  Also plotted are rotation
curves of the sample of LSB galaxies from vdH93.  The sample of vdH93
generally contains on average more massive LSB galaxies, as can be seen
in Fig.  \ref{Rot_HSB_LSB}. 

It is clear from Fig.  \ref{Rot_HSB_LSB} that the rotation curves of LSB
galaxies look similar to those of late-type ``normal'' galaxies, when
expressed in scale lengths.  We use the scale length of the optical disc
as this is the only easily observable length scale in a galaxy, which is
intrinsic to the galaxy, in sharp contrast with more arbitrary length
scales like isophotal radii or linear radii (see Section 7.1.1).  The
scale lengths of LSB galaxies are approximately twice as large as those
of HSB galaxies at a fixed maximum rotation velocity (ZHBM), so when
expressed in kpc, the rotation curves will rise more slowly to their
maximum velocity than those of HSB galaxies. 

We want to stress here that our rotation curves of LSB galaxies are
measured out to a comparable number of scale lengths as the average
rotation curve of a HSB galaxy (cf.  Fig.\ \ref{Rot_HSB_LSB}). 
Differences between the rotation curves can therefore not be caused by
sampling of different length scales due to the larger scale lengths of
LSB galaxies. 

\begin{figure}
\epsfxsize=8.5cm
\hfil\epsfbox{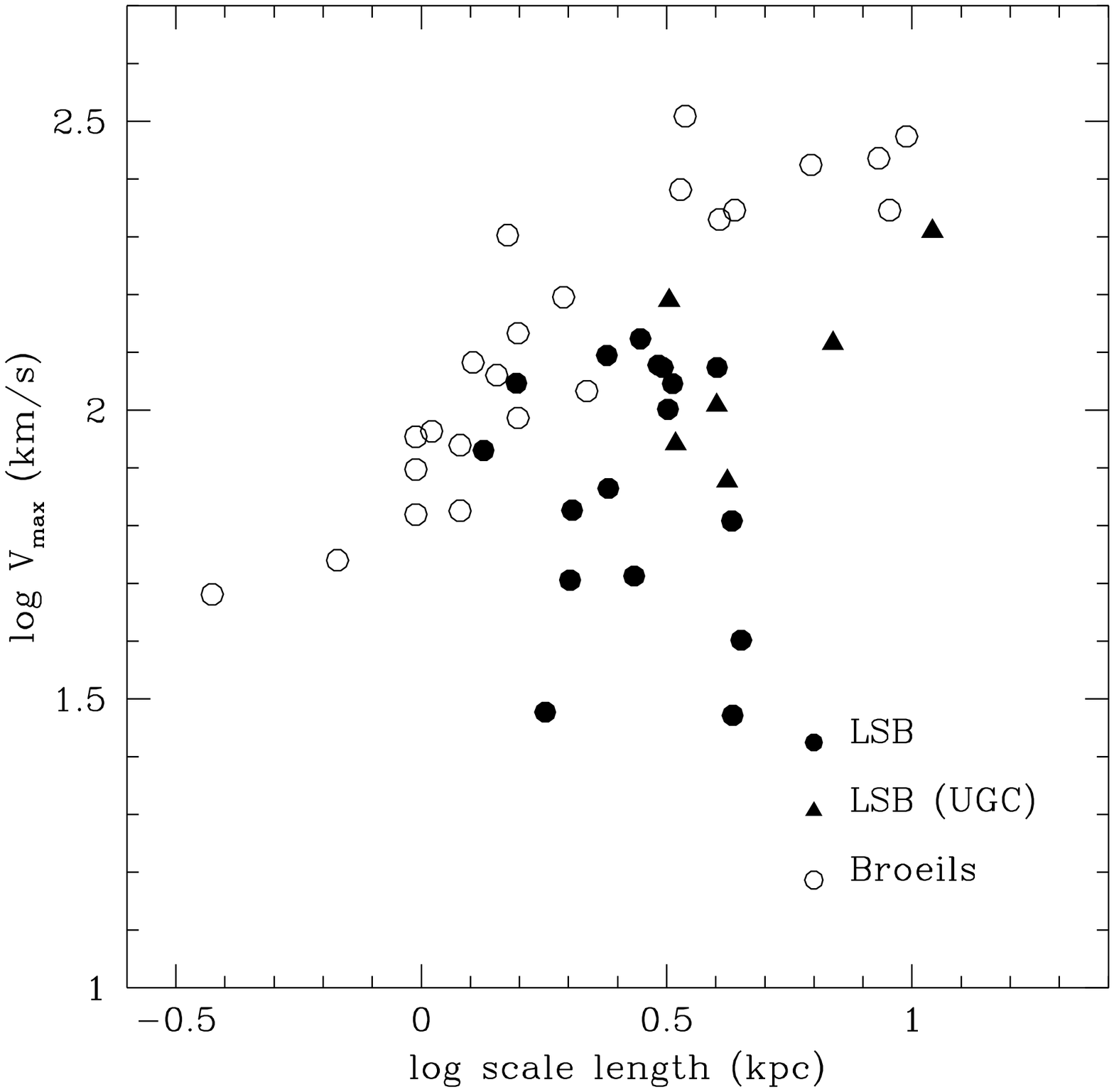}\hfil
\caption{Maximum rotation velocities plotted versus scale lengths. Open
symbols are all galaxies from Broeils (1992); filled circles are LSB
galaxies from our sample; filled triangles are LSB galaxies from vdH93.
 It is clear that LSB galaxies have larger scale lengths at a
fixed maximum rotation velocity.}
\label{Vmax_h}
\end{figure}

The larger values for scale lengths of LSB galaxies are
shown in Fig.  \ref{Vmax_h}, where the scale lengths in the HSB and LSB
samples are shown versus their corresponding maximum rotation
velocities.  The points that represent the LSB galaxies in general are located
at larger scale lengths than those of HSB galaxies at the same velocity. 
It is striking that the LSB galaxies only occupy the part of the figure
with $V_{\rm max} < 200$ km~s$^{-1}$.  Our sample therefore does not contain
the equivalents of the large HSB galaxies, that have been so well
studied in H{\sc i}.  We must probably find these large LSB galaxies
among the likes of UGC 6614: the Malin-1 cousins (Sprayberry et al. 1995a). 
 
\subsection{Compactness}

The outer parts of rotation curves are usually assumed to be determined
by the dark matter halos that surround galaxies (see van Albada and
Sancisi 1986).  The fact that LSB rotation curves rise more slowly than
those of HSB galaxies with the same $V_{\rm max}$ implies that not only
the light but also the matter in LSB galaxies is distributed more
diffusely.  A naive extrapolation would then suggest that this might
also apply to the dark matter, but an answer to this question will have
to wait for full disc-halo decompositions of the rotation curves, which
we will discuss  more
extensively in a forthcoming paper (de Blok \& McGaugh, 1996a). 

If the differences in surface density are the fundamental
criterion that distinguishes HSB galaxies from LSB galaxies, one would
like to have a quantitative measure for it.  We will therefore
use the compactness or the value of the characteristic mass surface
density $\sigma \propto V_{\rm max}^2 / h$, where $h$ is the exponential
scale length expressed in linear units.  It is easy to see that
combining the equations $M \propto V^2 h$ and $M \propto \sigma h^2$
leads to the above result.  Here $M$ is the dynamical mass within a
fixed number of scale lengths. 

We will now compare the compactnesses with the slopes of the rotation
curves.  Casertano and van Gorkom (1991), Persic \& Salluci (1991) and
Broeils (1992) already pointed out that there is a relation between the
outer slope of the rotation curve and the maximum rotation velocity. 
Low-luminosity dwarfs rotate slowly, and have rising rotation curves at
the edge of the observed H{\sc i} disc.  Compact high-luminosity
galaxies (with high rotation velocities) have declining rotation curves. 
As compactness in our definition depends on the maximum rotation
velocity this implies a relation between the compactness of a galaxy and
the slope of its rotation curve. 

Following Casertano and van Gorkom (1991) and Broeils (1992) we define
the slope as the slope of a straight-line fit to the logarithmic
rotation curve (i.e., $d \log V / d \log R$, as this ratio is distance-independent),
between an inner radius, defined as two-thirds of the optical radius
$R_{25}$, and the last measured point of the rotation curve.  The
decreasing $R_{25}$ towards later types and LSB galaxies does not affect
the value of the slope much. 

\begin{figure} 
\epsfxsize=8.5cm
\hfil\epsfbox{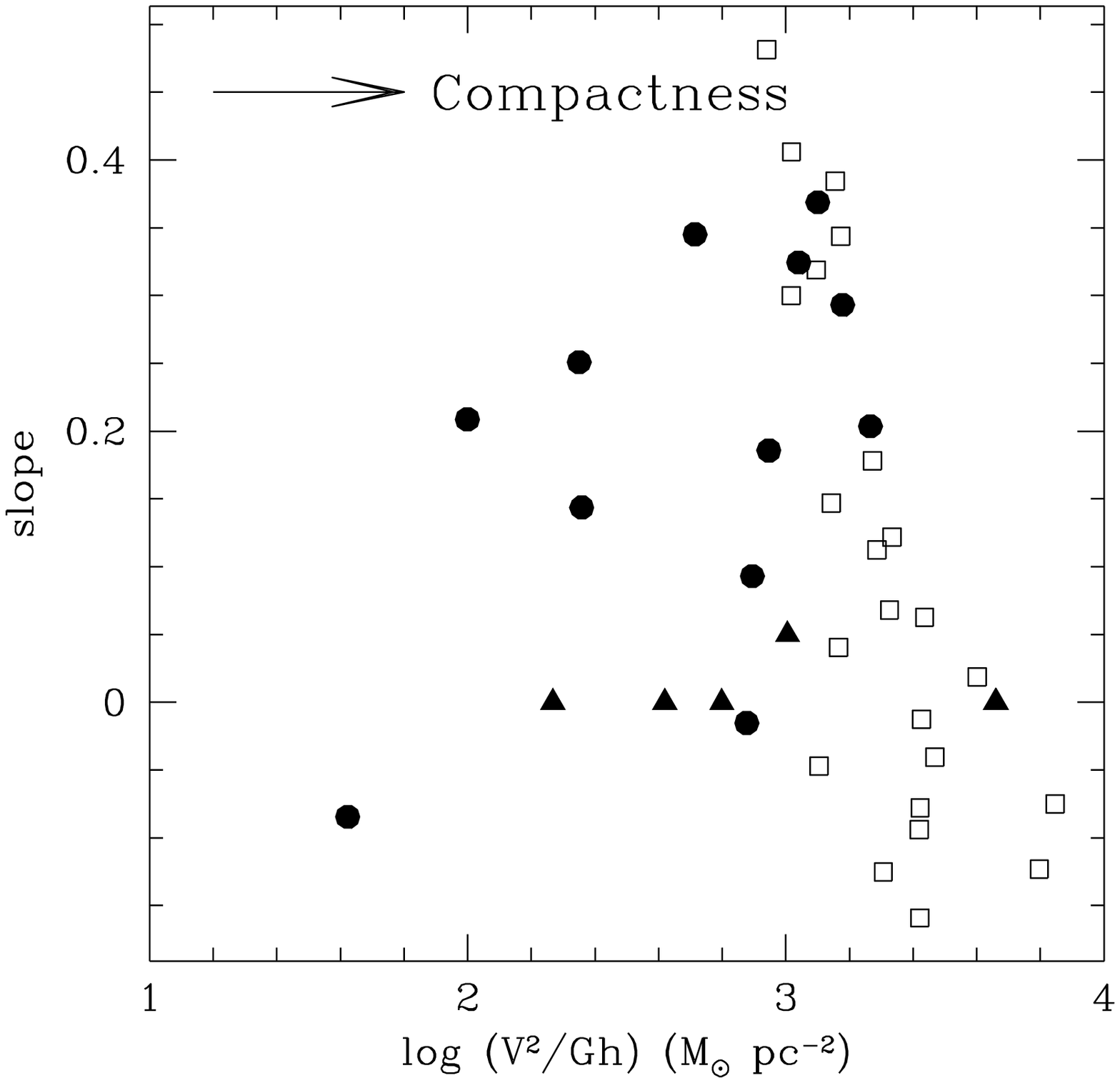}
\hfil 
 \caption{The compactness or mass surface density of HSB and LSB
galaxies.  It is clear that although some LSB galaxies have comparable
compactness as HSB galaxies, some have compactnesses that are a factor
of $\sim$10 smaller.  The open squares denote the galaxies from Broeils
(1992), while the filled circles represent the LSB galaxies from this
work.  Filled triangles are LSB galaxies from vdH93.}
\label{Compactness} \end{figure}

Figure \ref{Compactness} shows the comparison of compactness and slope
for HSB and LSB galaxies.  The rotation curves of the LSB galaxies have
positive slopes in the outer parts, but not more positive than those of
classical HSB galaxies.  In fact the only thing that distinguishes the
LSB galaxies from the HSB galaxies is that their ``compactness'' or mass
surface density is significantly lower than that of HSB galaxies with
comparable slope.  It is important to note that this surface density
includes all mass, light and dark matter. 

\begin{figure}
\epsfxsize=8.5cm
\hfil\epsfbox{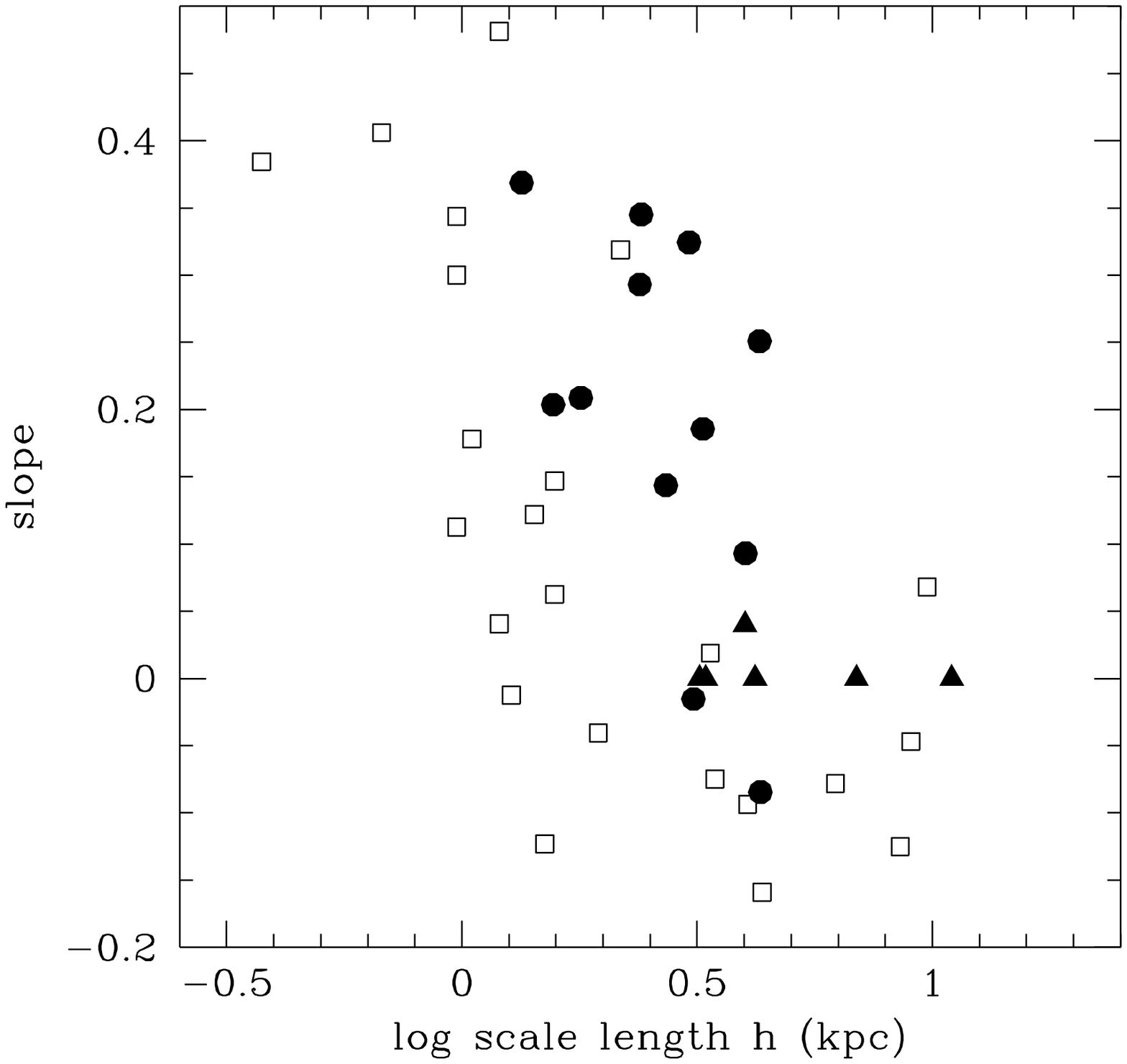}\hfil
\caption{The distribution of logarithmic slope $d \log V / d \log R$ of
the rotation curves of HSB and LSB galaxies with scale length. The open
dots denote the galaxies from Broeils (1992), while the filled circles
represent the LSB galaxies from this work. Filled triangles are LSB
galaxies from vdH93.}
\label{Slope_scale}
\end{figure}

We can also see that the trend Broeils (1992) has found of slope with
scale length has disappeared for scale lengths smaller than 5 kpc (see
Fig.  \ref{Slope_scale}).  At scale lengths smaller than 5 kpc the whole
range of slopes is covered.  At scale lengths larger than 5 kpc there
seem to be virtually no galaxies with positive slopes.  Most of the
galaxies plotted there are galaxies of types Sc and earlier.  However,
the presence of giant LSB galaxies at scale lengths between 8 and 10 kpc
suggests that the area between 5 and 8 kpc is only empty because of
selection effects.  Galaxies with identical scale lengths can have
rotation curves with different slopes.  There does, however, seem to be
a tendency for smaller galaxies to show a larger range of (mostly
positive) slopes than the larger galaxies. 

The rotation velocity, the shape of the rotation curve, the mass surface
density and the surface brightness are therefore all related to the
compactness of a galaxy.  This strongly suggests that the mass is not
the only parameter that determines the evolution of a galaxy.  {\it The
``compactness'' of a galaxy is also an important parameter for the
evolution of a galaxy}. 

\section{Mass to light ratios}

\subsection{Relation of $M_{\rm HI}/L$ with central surface brightness}
A tracer for the state of evolution of galaxies is the amount of H{\sc
i} gas that is present with respect to the amount of light: $M_{\rm
HI}/L_B$.  In order to get a good picture of how this ratio changes with
e.g.  luminosity or surface brightness one needs a sample of undisturbed
spirals of as large a range in central surface brightness and luminosity
as possible.  The sample of de Jong (1995) is a  good sample for
this purpose. 

We have taken the luminosities and H{\sc i} masses of this sample of 86
face-on spiral galaxies with accurate photometry and spanning a large range of
surface brightness, and calculated the $M_{\rm
HI}/L_B$ of the galaxies in the sample. The H{\sc i} data was taken from
the literature (RC3).  
\begin{figure}
\epsfxsize=8.5cm
\hfil\epsfbox{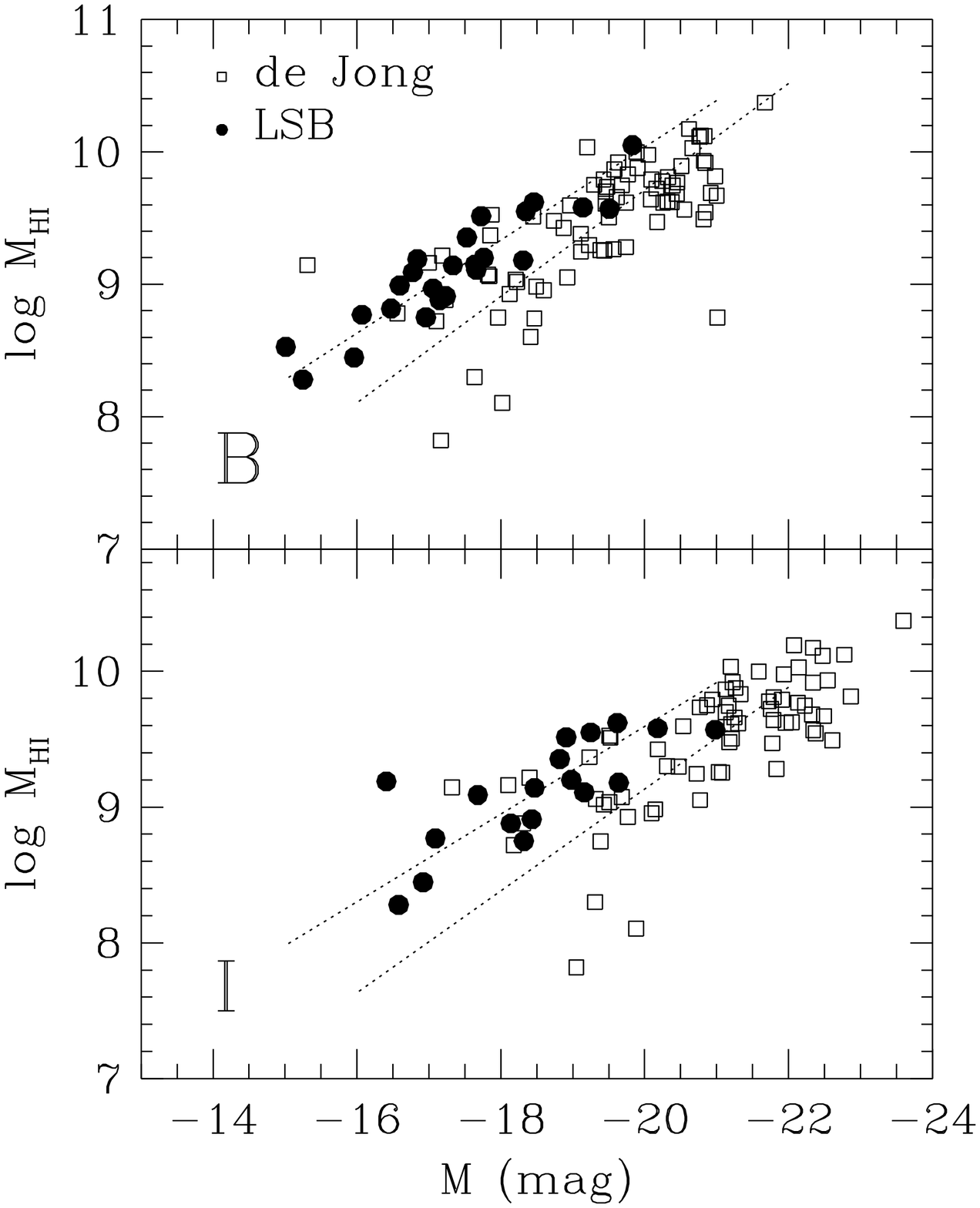}\hfil
\caption{The H{\sc i} masses of the galaxies in the sample of de Jong
and our sample plotted versus their luminosities. The top panel contains
$B$-band data. The bottom panel $I$-band data. The two 
lines are double least-square fits made to both samples 
separately. There is some overlap between the two samples as de Jong's sample
also contains a number of LSB galaxies. At a fixed luminosity LSB galaxies contain about 3 times
more H{\sc i}. }
\label{Bband}
\end{figure}

The top panel in Fig.  \ref{Bband} shows the H{\sc i} masses of the
galaxies in the sample plotted against their $B$-band luminosities.  It
is clear that LSB galaxies have more H{\sc i} at a fixed
luminosity than HSB galaxies.  To show that this effect is real and not
e.g.  caused by colour effects (as LSB galaxies are bluer), 
we have re-plotted the data in the bottom
panel of Fig.  \ref{Bband}, this time using the $I$-magnitudes.  This
shows that over a large range in luminosity LSB galaxies contain more
atomic gas per solar luminosity than their HSB counterparts. 
This may indicate that either LSB galaxies are really
gas-rich (i.e.  have a high gas-fraction $M_{\rm gas}/M_{\rm stars}$), or that
the importance of the (non-observed) molecular gas increases towards
higher surface brightnesses (so that $M_{\rm gas}/M_{\rm stars}$ might still have
similar values for all surface brightnesses).  However, in order to
explain the trend assuming the latter possibility, there needs to be
$\sim 10$ times more molecular gas than atomic gas in the highest surface
brightness galaxies (assuming that LSB galaxies have negligible amounts
of molecular gas, which seems to be supported by
observations\footnote{Recently Wilson (1995) has measured CO/H$_2$
conversion factors as a function of metallicity $Z$.  At $Z = 0.1
Z_{\sun}$ -- a typical value for the abundances in LSB galaxies -- the
factor increased by a factor of $\sim 5$, with respect to the Galactic
conversion factor.  This is not enough to allow for much H$_2$ in LSB
galaxies, based on the upper limits measured by Schombert et al.  (1990)
and de Blok (in prep.).}).
These large amounts are not consistent with determinations of
molecular gas masses in early-type HSB galaxies (Young \& Knezek 1989). 

The ratio $M_{\rm HI}/L$ is therefore larger for LSB galaxies, but as
Fig.  6 shows, both HSB and LSB galaxies have the same ratio
$\sigma_{\rm HI}/\mu_B$.  As LSB galaxies do form (and have formed)
fewer stars than HSB galaxies, but have higher gas fractions, the
conclusion must be that LSB galaxies have lower H{\sc i}
surface densities.  This means that the H{\sc i} disks {\it must} be more
extended in order to still reach their measured total H{\sc i} masses,
which are quite similar to those of HSB galaxies.  Clear 
differences  therefore exist in galaxies of different surface brightness.

\begin{figure}
\epsfxsize=8.5cm
\hfil\epsfbox{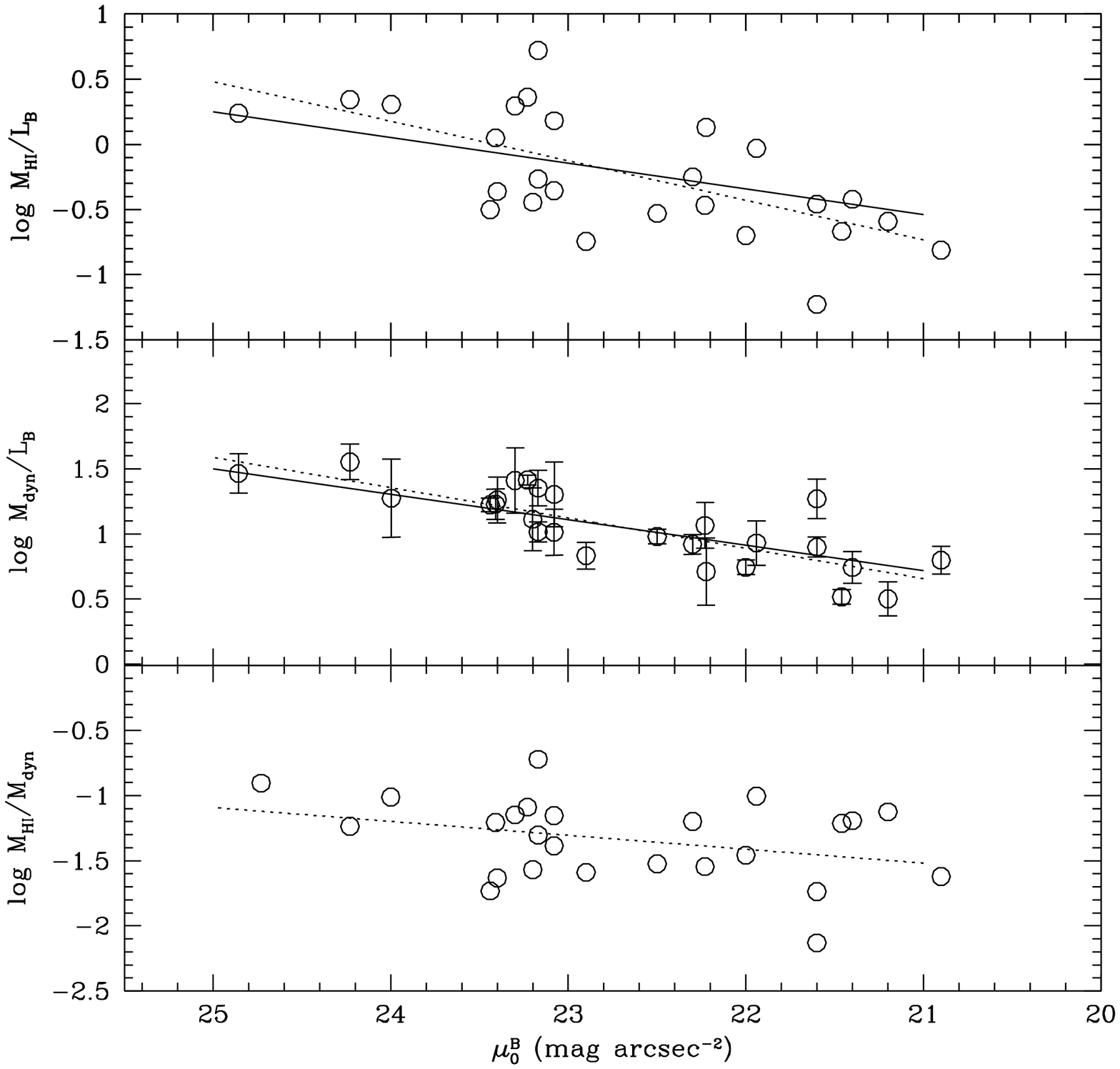}\hfil
 \caption{Various mass-to-light ratios for a sample of spiral galaxies
of different surface brightnesses.  The sample of Broeils (1992) was
used, together with LSB galaxies from this work with well-determined 
inclinations.  See Table A3 for identifications.  The top panel shows
the ratio of hydrogen mass to luminosity.  The full line the theoretical
expectation from the $\Upsilon\Sigma$-relation $(\Sigma\Upsilon^2 = \rm
constant$, where $\mu \propto -2.5 \log \Sigma$), normalized to go
through the data points.  The dotted line is the least-squares fit to
the data.  The middle panel shows the ratio of total mass within 4$h$ to
luminosity.  The full line is again the $\Upsilon\Sigma$-relation, and
the dotted line a least-squares fit to the data.  The bottom panel shows
the ratio of hydrogen mass and total mass within 4$h$.  The dotted line
is a least-squares fit to the data points.  }
 \label{large}
\end{figure}
 
\subsection{Relation of $M_{\rm dyn}/L$ with central surface brightness}
 Gas and stars form only the visible part of galaxies.  In order to find
possible relations of surface brightness with the total mass of galaxies
one needs to investigate the behavior of $M_{\rm dyn}/L_B$.  Here we
will only discuss this parameter in a very global way.  An extensive
discussion of $M_{\rm dyn}/L_B$, using disc-halo decompositions of the
rotation curves will be the subject of a separate paper.  We will merely
note that a strong relation of $M_{\rm dyn}/L_B$ with surface brightness
exists.  This is shown in the middle panel of Fig.  \ref {large} where
the mass to light ratio $M_{\rm dyn}/L_B$ is plotted as a function of
central surface brightness.  The HSB galaxies are from the sample of
Broeils (1992), the LSB galaxies are from our sample.  We have adopted
four scale lengths as a consistent definition (MB94, BHB95, ZHBM) of the
optical radius, where we determine both the mass and the light.  This
then is the $M/L$ within the adopted ``optical radius'' $R_{\rm opt} =
4h$.  We of course cannot measure the {\it total} mass, just that
enclosed by the light.  Note that these mass measurements are sensitive
to the inclination corrections.  It is thus important that the
inclination be accurately measured, and we have retained only galaxies
which have adequate inclination determinations.  Errorbars in Fig.  \ref
{large} are based on a pessimistic error estimate of 6 degrees.  See
Table A3 for the identifications of the galaxies plotted. 

 The derived mass is also sensitive to the measured value of the
asymptotic velocity itself, so we have used only those galaxies for
which this can be taken directly from a full rotation curve.  A strong
correlation is apparent: this is the mass-to-light-ratio ($\Upsilon$) -- surface
brightness ($\Sigma$) relation from ZHBM.  At the highest surface
brightnesses, the mass to light ratio is similar to that expected from
stellar population models.  At the faintest surface brightnesses, it has
increased by a factor of ten, indicating increasing dark matter
domination within the optical disc as surface brightness decreases,
maybe in combination with a systematic change in the stellar population. 
The slope of the full line in Fig.  \ref{large} was forced to equal the
value required by the $\Upsilon\Sigma$-relation and has been shifted to
go through the data points.  The dotted line is the actual least-squares
fit to the data points.  These two fits are identical
within the errors. 

\subsection{Hydrogen and total masses} In the top panel of Fig. 
\ref{large} we have plotted $M_{\rm HI}/L_B$ for the same galaxies as a
function of central surface brightness.  It is striking that this
bears a strong resemblance to the
$\Upsilon\Sigma$-relation.  The full line shown in the top panel of Fig. 
\ref{large} is the $\Upsilon\Sigma$-relation.  The dotted line is again the actual
least-squares fit to the data.  The $\Upsilon\Sigma$-relation, which is
a relation between dynamical mass and surface brightness, is therefore
also a good description for the relation between gas mass and surface
brightness.

The relation shown in the top panel of Fig.  \ref{large} is independent
of the dynamics of the galaxies, while the $\Upsilon\Sigma$-relation is
not.  This strong relation of $M_{\rm HI}/L_B$ strongly suggests an
evolutionary solution to the $\Upsilon\Sigma$-relation, as argued by ZHBM. 
This could, for
example, be in the form of H{\sc i} as a tracer of dark matter (Bosma
1981), or in the form of dark matter as cold gas (Pfenniger et al 1994).
A possible problem for an evolutionary solution, however, is that the
large values for $M_{\rm dyn}/L$ derived for LSB galaxies imply that
the contribution of baryons is not important. 

The slight difference between the slopes in the top and middle panels of
Fig.  \ref{large} introduces a small residual trend of $M_{\rm HI}/M_T$
with surface brightness.  This is presented in the bottom panel of Fig. 
\ref{large}.  The dotted line is a least-squares fit to the data.  This
trend may be caused by the increasing importance of atomic gas towards
lower surface central surface brightnesses. 
Rotation curve decompositions are needed to solve this puzzle.
However, real variations in $M_{\rm HI}/M_T$ would make evolutionary
scenarios less attractive.

We therefore conclude that LSB galaxies are more gas-rich than HSB
galaxies, implying that they have systematically higher gas-fractions. 
This shows again clearly that evolutionary differences exist between galaxies
of different surface brightnesses.  

\section{Concluding remarks}

In this paper we have presented the H{\sc i} properties of a sample of
19 late-type LSB galaxies, representative of the LSB galaxies in the
SBSM catalogue, determined from observations with the {\sc vla} and {\sc wsrt}. 

We have shown that these late-type LSB galaxies  have low H{\sc i} surface
densities, both average and local values are about a factor of $\sim 3$
lower than those in comparable late-type HSB galaxies. 

By comparing local surface brightnesses and H{\sc i} surface densities
from galaxies over a large range in surface brightness, we show
that both HSB and LSB galaxies follow the same surface
brightness--atomic gas surface density trend.  LSB galaxies may in
that respect resemble the low surface brightness outer parts of HSB
galaxies.  Other evidence (e.g.  colours and metallicities) suggests
that they may be less evolved.  Local surface density seems to control
the rate of evolution. 

Rotation curves rise to maximum velocities between 50 and 120 km
s$^{-1}$, and usually have not flattened off in the outermost point. 
They rise more slowly towards their asymptotic value than those of
comparable HSB galaxies, suggesting an overall lower mass surface
density (see also de Blok \& McGaugh 1996b): one might speculate that
even the dark matter of LSB galaxies may have a lower surface density. 
A confirmation of this will have to wait for full disc-halo
decomposition analysis of the rotation curves  (de Blok \& McGaugh 1996a).

The late-type LSB galaxies in our sample have higher gas fractions than
HSB galaxies, and establish a relation between $M_{\rm HI}/L_B$ and
central surface brightness.  A similar relation between $M_{\rm
dyn}/L_B$ and central surface brightness also exists.  This confirms the
$\Upsilon\Sigma$-relation found by ZHBM.  All of this may point at an
evolutionary solution for the $\Upsilon\Sigma$-relation, but the high
values for $M_{\rm dyn}/L$ argue against this.  This remains a major
puzzle. 

Evolution, compactness, surface density and surface brightness are
intimately related.  The LSB galaxies we have investigated have low
baryonic surface densities and are in all probability slowly
evolving systems. 

\section*{Acknowledgments}

We thank Adrick Broeils for making some of his tables available in
electronic form and Adwin Boogert for doing the observations at the
Dutch Telescope at La Silla.  We also like to thank Renzo Sancisi and 
Rob Swaters for many
useful discussions and constructive suggestions.  dB would like to thank
the Leidsch Kerkhoven-Bosscha Fonds for travel grants during the writing
of this paper. We thank Jon Davies for his comments on an earlier
version of this paper.

\appendix
\section{Additional optical observations}
\subsection{Observations and reduction}
\subsubsection{Jacobus Kapteyn Telescope}
CCD imaging for a couple of galaxies in our sample, for which previously
no optical information was available (F571-8, F571-V2, F574-1, F579-V1,
F583-1 and  F583-4), was obtained at the 1 meter
Jacobus Kapteyn Telescope\footnote{The Jacobus Kapteyn Telescope is
operated by the Royal Greenwich Observatory at the Observatorio del Roque de los Muchachos of the
Instituto de Astrof\'\i sica de Canarias with financial support from the
PPARC (UK) and NWO (NL).}
at La Palma, during two nights in March 1995.

Conditions were variable, with an average seeing of 2 to 3
arcsec. 
$R$-band imaging was obtained using the 1124$^2$ TEK chip at the JKT,
which was binned to 562$^2$ due to the poor seeing conditions. The pixel
size therefore became 0.662 arcsec. Images were reduced using standard
procedures with the {\sc iraf} reduction package.
See Table A1 for specific details on the observations. See BHB95
for a complete description of the reduction procedures.

The images were flux-calibrated using standard stars from the list of
Landolt (1992).  Due to the variable conditions the photometric
solution, based on 15 standard star exposures, is formally only
accurate to 0.2 magnitudes, but as will be shown later, is likely to be
much better than this. 

Many images showed a gradient across the sky background. This gradient
was removed during the process of sky subtraction, but deviations from
the mean sky background resulting from this gradient were later used to
estimate uncertainties in the surface brightness of the galaxies.
This procedure is also described in BHB95.

\subsubsection{ESO Dutch Telescope}
$BRI$ CCD imaging of F583-1 was obtained at the 90 cm Dutch Telescope at
ESO, Chile, during the night of April 14-15 1994. Conditions were 
photometric, with an average seeing of 1 arcsec. Imaging was obtained
using a 512$^2$ TEK chip, with a pixel size of 0.457 arcsec. Images were
reduced using standard procedures (see above). Calibration was performed
using Landolt (1992) standard stars. Photometric solutions had an
accuracy of 0.03 mag in $B$, $R$ and $I$. See Table A1 for a journal of
the observations.

\begin{table}
\begin{minipage}{100 mm}
\caption[]{Journal of observations}
\begin{tabular}{llll}
\hline
Name    & date   & filter & exposure \\
(1)&(2)&(3)&(4)\\
\hline
{\bf La Palma}&&&\\
F563-1 & 	Mar 7 '95  &R&	4$\times$900\\
       &	Mar 8  &R&	2$\times$900\\
F571-8 & 	Mar 2  &R&	4$\times$900, 600 \\
F571-V2& 	Mar 2  &R&	2$\times$900, 600 \\
F574-1 & 	Mar 2  &R&        4$\times$900\\
F579-V1 &       Mar 2  &R&        3$\times$900\\
F583-1 &        Mar 7  &R&        3$\times$900\\
F583-4 &  	Mar 8  &R&	2$\times$900\\
\hline
{\bf La Silla}&&&\\
F583-1 & 	Apr 14 '94 &B&	9$\times$300\\
	&		&R&	7$\times$300\\
	&		&I&	6$\times$300\\
\hline
\end{tabular}

(1) Name of the galaxy\\
(2) Date at the beginning of the night\\
(3) Photometric band\\
(4) Exposure time in seconds of individual observations\\
\end{minipage}
\end{table}

\subsection{Radial profiles} 
\subsubsection{Derivation}
For each galaxy ellipse fit were made to
the outer isophotes, and radial profiles were produced using the
inclinations and position angles thus derived.  The central surface
brightness and scale length of the disc were derived by using a
decomposition of the profile in an exponential disc and an exponential
bulge.  In practice the contribution of the bulge was for most of the
galaxies negligible. 

The optical structural parameters thus derived are given in Table A2,
along with the total magnitudes and isophotal radii. Note that the
central surface brightnesses are the measured values: no correction has
yet been made for inclination and Galactic foreground extinction.
Figure A1 presents the radial $R$-band profiles of F563-1, F574-1, F579-V1 and
F583-4.  

\begin{figure}
\epsfxsize=8.5cm
\hfil\epsfbox{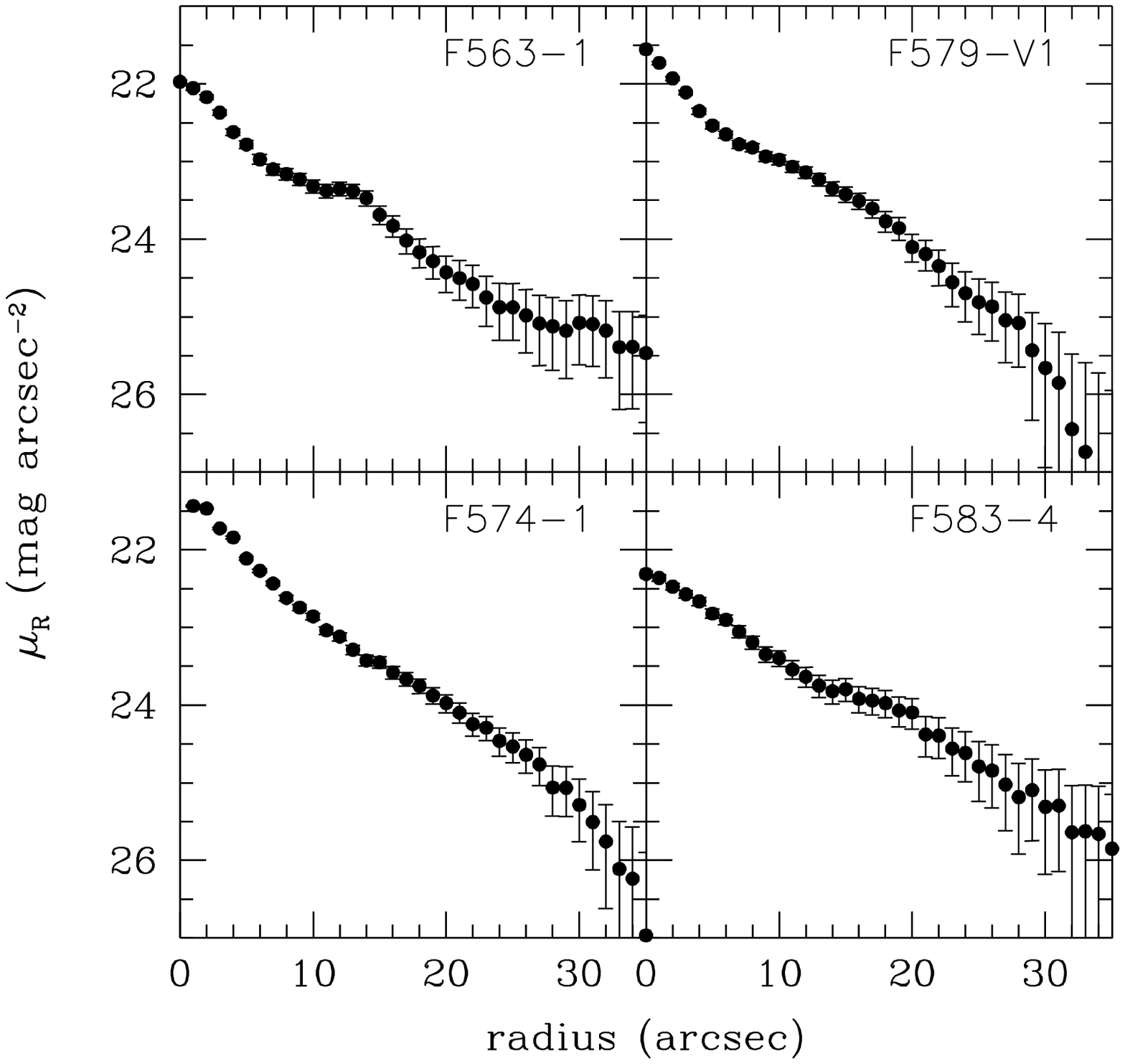}\hfil
\caption{Radial $R$-band profile of F563-1, F574-1, F579-V1 and F583-4.}
\label{JKTpanel}
\end{figure}

\subsubsection{Note on F563-1: Comparison with other data}
Comparison with BHB95 will show that the optical
inclination of F563-1 given here is substantially different from their
inclination. This is because the image in BHB95 suffered from a
modest amount of stray light and it was therefore not clear whether the
extended face-on like structure around F563-1 was real or not. The
images obtained at the JKT show that indeed this galaxy is surrounded by
a faint face-on-like disc. The old inclination in BHB95 was
determined using
the inner barlike structure, yielding a value of 50 degrees. The new
value of 25 degrees derived here, is in much closer agreement with the H{\sc i} data.

We have determined the radial profile of F563-1 from the new JKT imaging
data, using the inclination and position angle given in BHB95.
This is shown in Fig. A2, where we have also plotted their INT profile of
F563-1. The two profiles are clearly indistinguishable.
Only at very low surface brightnesses does the JKT profile diverge from
the INT profile. This is simply because the JKT imaging has a lower
signal to noise ratio than the INT imaging of BHB95.
This suggests that the surface brightnesses presented here are likely to
be more accurate than the uncertainty of 0.2 magnitudes which is given
by the formal photometric fit to the standard stars.

\begin{figure}
\epsfxsize=8.5cm
\hfil\epsfbox{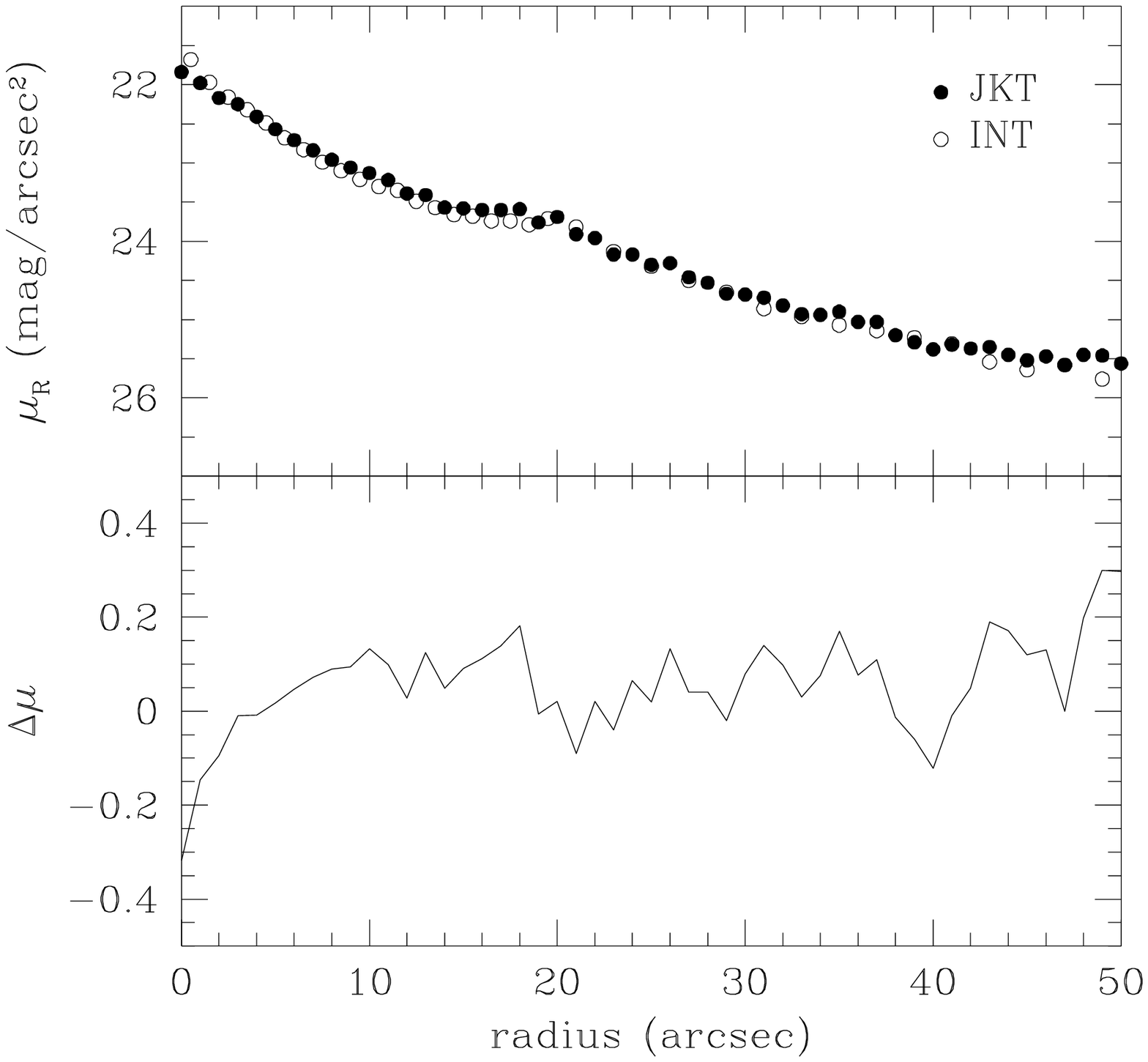}\hfil
\caption{Comparison of the radial profiles of F563-1, using the new data
presented here (closed dots) and the data from BHB95
(open dots). To produce the profiles the ellipse parameters from BHB95
have been used.}
\label{F5631comp}
\end{figure}

\begin{figure}
\epsfxsize=8.5cm
\hfil\epsfbox{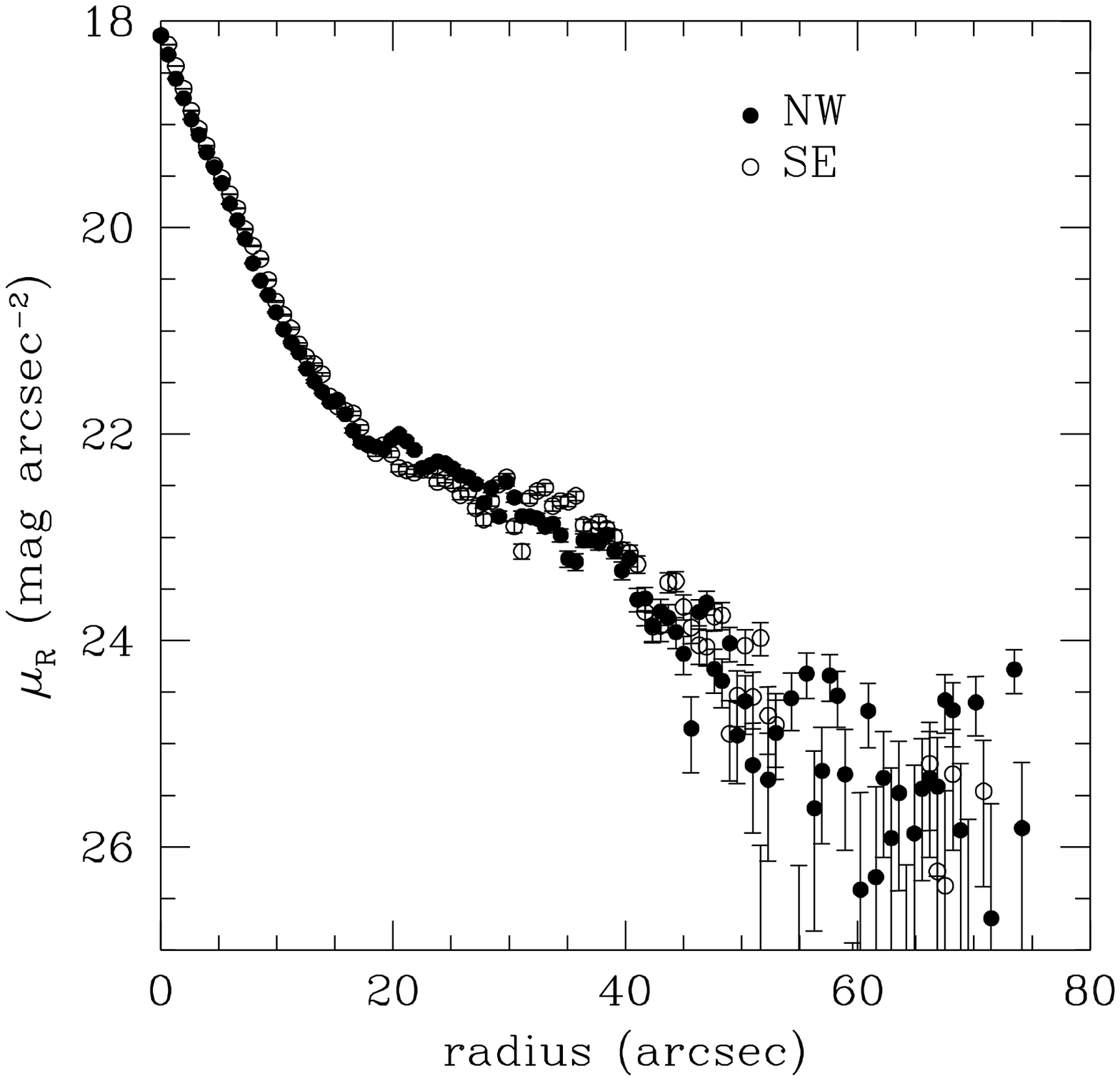}\hfil
\caption{Major axis $R$-band profile of F571-8. The profile has a width
perpendicular to the major axis of 1.5 arcsec. The closed circles denote
the major-axis light distribution of the NW side of the galaxy, the open
circles that of the SE side.}
\label{F5718comp}
\end{figure}

\subsubsection{Notes on F571-8: an edge-on LSB galaxy}

This galaxy is the only edge-on galaxy in our sample.  During the
initial sample selection the faint edge-on disc was not noticed, and
only showed up in the H{\sc i} and subsequent optical observations.  As
the usual ellipse fitting methods and integrating in rings to produce
the radial profile, do not work for edge-on galaxies, we have produced
the radial profile as follows.  A major axis cut of 2 arcsec width
(corresponding to one seeing FWHM) was made, and this profile was
subsequently decomposed into a bulge and disc component.  This profile is
shown in Fig. A3, where the profile is folded around the central
position.  An exponential bulge was fit to the inner brightness peak,
subtracted, and another exponential disc was fit to the residual
profile.  Structural parameters of these components can be found in
Table A2.  This galaxy is also special in the sense that it has a very
significant bulge, in contrast with all other galaxies in the sample. 
The magnitudes of the bulge and disc components, and the total galaxy,
were measured directly from the optical image by integrating the total
fluxes of the components. The structural parameters were therefore not
used in determining these quantities. Using these magnitudes yields a
bulge-to-disc ratio for F571-8 of $\sim 2$, which is  much larger
than those of the other galaxies in the sample where usually $B/D \ll 0.1$. 

The image shows no visible evidence for dust: there is no sign of an
equatorial dust lane. This is confirmed by the minor axis profiles: the
light distribution above and below the plane is identical for all radii.
If F571-8 were representative for the face-on LSB galaxies in this
sample, this would confirm the low dust contents for LSB galaxies
inferred from colours, observed Balmer decrements and CO fluxes.
However, the large bulge of this galaxy probably means that this galaxy
belongs in a different class of LSB galaxies. A more extensive study of
edge-on LSB galaxies should shed
some more light on this.

\subsubsection{Notes on F583-1: Comparison and colours}

Using the procedures described in BHB95 we have determined the radial
luminosity and colour profiles of F583-1. These are presented in Fig.
A4. The leftmost panels show the radial profiles in $B$, $R$ and $I$.
The errors represent the maximum errors caused by uncertainty in the
value of the sky background. Even though this uncertainty was 3 counts
typically, the effects on the values of the surface brightness can be
quite severe. Values for surface brightness and scale length are given
in Table A2. Comparison with other galaxies from BHB95 shows that this
galaxy is a typical LSB galaxy in all respects. The colours (shown in
top-right panel of Fig. A4) are also quite normal for LSB galaxies.
We find that in F583-1 the average (area weighted) value for $B-R$ is
0.95, which is slightly redder than the typical value of the sample of
BHB95, but still well within the range covered by LSB galaxies. The area
weighted value for $B-I$ is 1.35, which is almost the typical value
found for the sample of BHB95.

The two panels at bottom-right compare the present data with data
obtained with the JKT telescope (see Section A1.1).  It is clear that
despite the variable conditions during the JKT run, the quantities
derived from that data are usually accurate to within a tenth of a
magnitude. 

\begin{figure}
\epsfxsize=8.5cm
\hfil\epsfbox{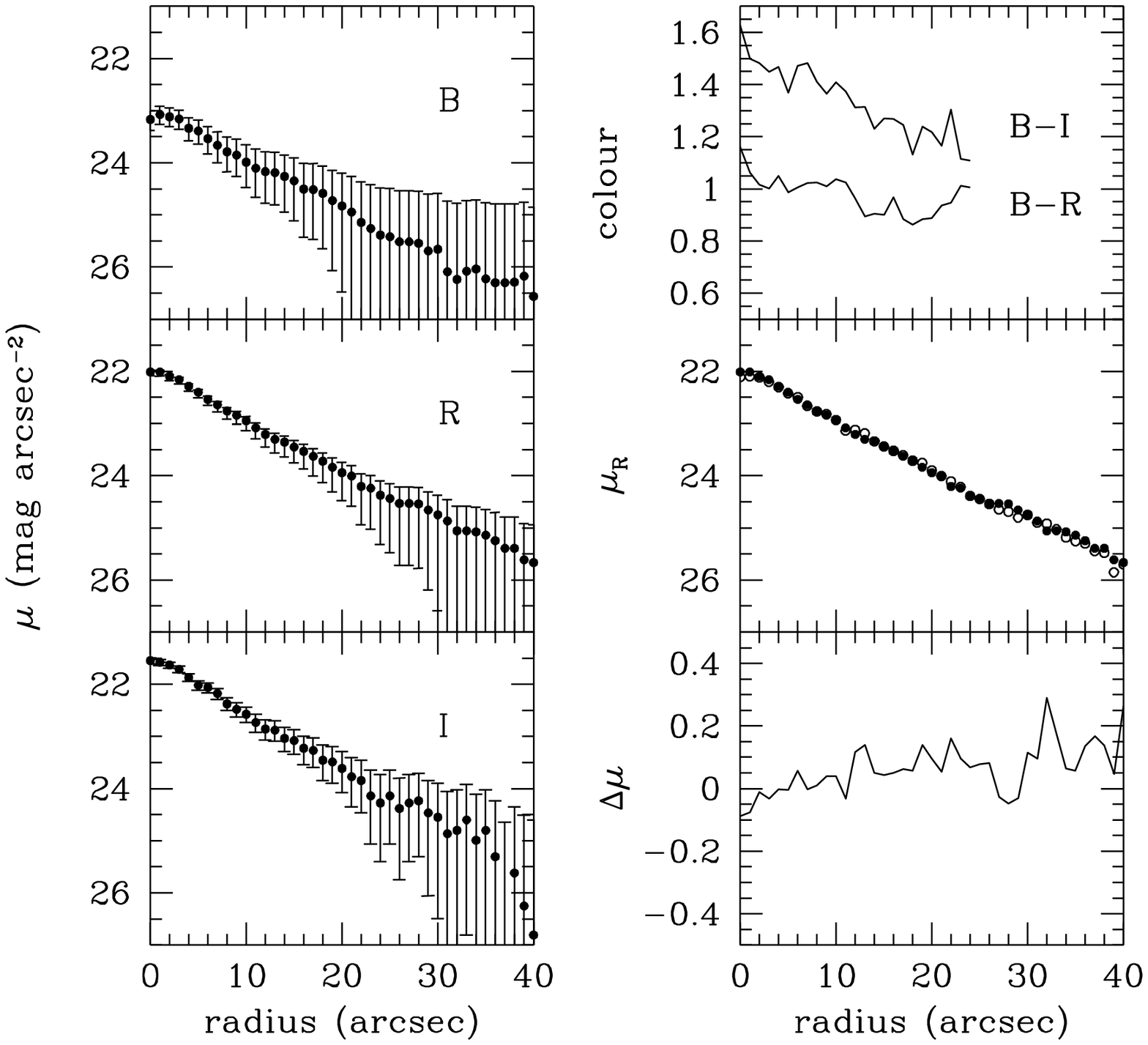}\hfil
\caption{Radial profiles of F583-1. Left row shows, from top to bottom,
the measured radial luminosity profiles in  the $B$, $R$ and $I$ bands
respectively. The right row shows in the top panel the $B-I$ and
$B-R$ radial colour profiles; the middle panel shows a comparison
between the $R$ band profiles as obtained with the JKT (open circles)
and the Dutch Telescope (filled circles); the bottom panel shows the
difference between these two profiles $\Delta \mu = \mu_{Dutch} - \mu_{JKT}$.}
\label{F5831pan}
\end{figure}

\subsection{Conversion to $B$-band}

In order to make this data directly comparable with the data from BHB95,
used in this paper to discuss the properties of LSB galaxies, we have
converted the $R$ surface brightnesses and magnitudes of F571-8, F574-1,
F579-V1 and F583-4 to the $B$-band using the median colours of the
sample of LSB galaxies discussed in BHB95. 

As the median $B-R$ colour of LSB galaxies is $+0.78$, we have assumed
that $\mu_B(0) = \mu_R(0) + 0.78$.  This value for $\mu_B(0)$ was
furthermore corrected for inclination, and these are then the values
given in Table 3. 

As is shown in BHB95, there is also a small change in scale length
between $B$ and $R$-band. However this change is much smaller than the
uncertainty introduced by the correction of the surface brightness, and
we have therefore assumed that $h_B = h_R$.

Total $R$-band magnitudes underwent a similar correction: as we have
assumed the scale length to be constant, we can compute $m_{T,B} =
m_{T,R} + 0.78$. From this one can compute values for  $L_B$ and $M_B$,
which are given in Table 3.

\begin{table*}
\begin{minipage}{120mm}
\caption[]{Parameters of the sample.}
\begin{tabular}{llrrrrrrrrrrr}
\hline
Name   &filter& $i$  & PA & $\mu_{\lambda}$ & $\alpha$ & $h$ & $m_T$ & $R_{25}$ & $m_{25}$ & $M_T$ &
$A_{\lambda}$ & Note \\ 
(1)&(2)&(3)&(4)&(5)&(6)&(7)&(8)&(9)&(10)&(11)&(12)&(13)\\
\hline
F563-1 &R& 25 & 168 & 22.55 & 13.0 & 2.1 & 15.1 & 26 & 16.3 & --17.6 & 0.04& \\
F571-8 &R& 90 & 167 &   *   &  *   &  *  & 15.0 & 33 & 15.0 & --17.8  & 0.00& $a$\\
\ disc && *  &  *  & 20.59 & 16.0 & 2.8 & 16.2 &  * &  *   & --16.6 &  * & $b$\\
\ bulge&& *  &  *  & 18.13 & 3.3 & 0.6 & 15.4 &  * &  *   & --17.4 &  * &$b$\\
F574-1 &R& 65 & 92  & 21.60 &  9.1 & 3.2 & 15.7 & 38 & 15.8 & --18.6 & 0.05& \\
F579-V1&R& 26 & 113 & 21.90 & 10.0 & 3.8 & 15.0 & 27 & 15.9 & --19.0 & 0.04& \\
F583-1 &B& 63 & 175 & 23.05 & 10.2 & 1.2 & 16.0 & 21 & 16.5 & --15.9 & 0.09\\
       &R& 63 & 175 & 22.01 & 9.8  & 1.1 & 15.1 & 32 & 15.3 & --16.8 & 0.06& \\
       &I& 63 & 175 & 21.38 & 9.1  & 1.1 & 14.6 & 33 & 14.8 & --17.1 & 0.04\\
F583-4 &R& 55 & 116 & 22.38 & 11.3 & 2.0 & 15.7 & 27 & 16.1 & --17.1 & 0.02& \\
\hline 
\end{tabular}\\
{\em Columns:}\\
(1) Name of the galaxy.\\
(2) Photometric band.\\
(3) Inclination as derived from our data. \\
(4) Position angle measured from N through E.\\
(5) Uncorrected central surface brightness of disc.\\
(6) Scale length $a$ in arcsec.\\
(7) Scale length $h$ in kpc, using distance from Table 3.\\
(8) Total magnitude of exponential disc.\\
(9) Radius in arcsec of 25 $R$-mag arcsec$^{-2}$ ellipse.\\
(10) Total magnitude $m_{25}$ within 25 $R$-mag arcsec$^{-2}$ ellipse.\\
(11) Absolute magnitude, using column 7 and Table 3.\\
(12) Applied extinction correction in $R$. \\
(13) Notes: {\bf a)} Magnitudes are determined from the total image, by
integrating the flux. $R_{25}$ has been measured in the major axis cut
presented in Fig. A3. {\bf b)} The magnitudes of
the bulge and disc components are determined directly from the image.
The structural parameters were determined from the major-axis cut.\\

\end{minipage}
\end{table*}

The corrections made to the values derived for F571-8 are of a different
nature. In converting the edge-on central surface brightness of the disc, as
measured from the major axis profile, to a face-on surface brightness,
we have assumed the disc to be completely optically thin. This is
probably an oversimplification, however, evidence that LSB galaxies may
not contain as much dust as normal HSB galaxies, also invalidates using the
standard corrections for internal extinction. In the case of zero
internal extinction the face-on central surface brightness is simply
half that as measured edge-on (0.75 magnitudes fainter). To this face-on
value we also added the colour correction of 0.78 magnitudes, yielding
the value given in Table 3.
The same assumptions were made in determining the luminosity of the
disc: optically thin, and a colour correction of 0.78 magnitudes.

\subsection{Identification of galaxies in Fig. 13}

\begin{table}
\begin{minipage}{70mm}
\caption[]{Identification of galaxies in Fig. \ref{large}}
\begin{tabular}{llll}
\hline
Name&Ref&Name&Ref\\
(1)&(2)&(3)&(4)\\
\hline
DDO 154 &  1,2	 & NGC 5033 & 3,9 \\ 
DDO 168 &  3	 & NGC 5585 & 13 \\
NGC 55 	&  4     & NGC 6674 & 3 \\
NGC 247 &  5,6	 & NGC 7331 & 3,9 \\
NGC 300 &  7,6	 & UGC 2259 & 14,9  \\ 
NGC 801 &  3,8	 & UGC 2885 & 15,8 \\
NGC 1560&  3	 & F568-3   & 16,17\\
NGC 2403&  9,10	 & F568-V1  &16,17\\
NGC 2841&  9,10	 & F571-V1  &16,17\\
NGC 2903&  9,10	 & F563-V2  &18,17\\
NGC 2998&  3,8	 & UGC 128  &19\\
NGC 3109&  11,10 & UGC 6614 &19\\
NGC 3198&  12,10 & & \\
\hline 
\end{tabular}\\
{\em Columns:}\\
(1),(3) Name of the galaxy.\\
(2),(4) Source of the data. \\
Sources:\\
{[1]} Carignan \& Freeman (1985)         \\
{[2]} Carignan \& Beaulieu (1989)          \\
{[3]} Broeils (1992)                       \\
{[4]} Puche, Carignan \& Wainscoat (1991)  \\
{[5]} Carignan \& Puche (1990)             \\
{[6]} Carignan (1985)\\
{[7]} Puche, Carignan \& Bosma (1990)\\
{[8]} Kent (1986)\\
{[9]}  Begeman (1987)\\
{[10]}  Kent (1987)\\
{[11]} Jobin \& Carignan (1990)\\
{[12]} Begeman (1989)\\
{[13]} C\^ ot\' e, Carignan \& Sancisi (1991)\\
{[14]} Carignan, Sancisi \& van Albada (1988)\\
{[15]} Roelfsema \& Allen (1985)\\
{[16]} BHB95\\
{[17]} this work\\
{[18]} MB94\\
{[19]} vdH94\\
\end{minipage}
\end{table}

Table A3 contains the identification of the galaxies plotted in Fig. 
\ref{large}.  Also given are the sources for the data.  We retrieved the
surface brightnesses from the original papers (see notes to Table
A3 for a complete list), and where necessary converted to the $B$-band. 


\begin{thebibliography}{}
\bibitem[]{}
Allen, R.J., Shu, F.H., 1979, ApJ, 227, 67
\bibitem[]{}
Begeman K.G., 1987, PhD thesis, University of Groningen
\bibitem[]{}
Begeman K.G., 1989, A\&A, 223, 47
\bibitem[]{}
Bergvall N., R\" onnback J., 1994, A\&AS, 108, 193
\bibitem[]{}
Bos A., Raimond E., van Someren Greve H.W., 1981, A\&A 98, 251 
\bibitem[]{}
Bosma A., 1981, AJ, 86, 1825
\bibitem[]{}
Bothun, G.D., Schombert, J.M.,  Impey, C.D., Sprayberry, D.,  McGaugh, S.S.,
1993, AJ, 106, 530
\bibitem[]{}
Broeils A.H., 1992, PhD thesis, University of Groningen
\bibitem[]{}
Carignan C., 1985, ApJSup, 58, 107
\bibitem[]{} 
Carignan C., Beaulieu, S., 1989, ApJ, 347, 760
\bibitem[]{}
Carignan C., Freeman, K.C., 1988, ApJLett, 332, L33
\bibitem[]{}
Carignan C., Puche, D., 1990, AJ, 100, 641
\bibitem[]{}
Carignan C., Sancisi R., van Albada T.S., 1988, AJ, 95, 37
\bibitem[]{}
Casertano S., van Gorkom J.H., 1991, AJ, 101, 1231
\bibitem[]{}
Cayatte V., Kotanyl C., Balkowski C., van Gorkom J. H., 1994, AJ, 107, 1003
\bibitem[]{}
C\^ ot\' e S., Carignan C., Sancisi R., 1991, AJ, 102, 904
\bibitem[]{}
Dalcanton J.J., Spergel D.N., Summers F.J., 1995, preprint
\bibitem[]{}
Davies, J. I., Disney, M. J., Phillips, S., Boyle, B. J., Xouch, W. J,
1994, MNRAS, 269, 349
\bibitem[]{}
Davies, J. I., Phillips, S., Disney, M. J., 1988, MNRAS, 231, 69p
\bibitem[]{}
de Blok W.J.G., McGaugh, S.S., 1996a, in preparation
\bibitem[]{}
de Blok W.J.G., McGaugh, S.S., 1996b, ApJL, in press
\bibitem[]{}
de Blok W.J.G., van der Hulst J.M., Bothun G.D., 1995, MNRAS, 274, 235 (BHB95)
\bibitem[]{}
de Jong R.S., 1995, PhD thesis, University of Groningen
\bibitem[]{}
de Jong R.S., 1996, A\&A, in press
\bibitem[]{}
de Vaucouleurs A., de Vaucouleurs G., Buta R.J., Corwin Jr. H.G.,
Fouqu\'e P.,  Paturel G., 1991, { Third Reference Catalogue of Bright
Galaxies (RC3)}, Springer-Verlag, New York
\bibitem[]{}
Flores, R. A., Primack, J. R., 1994, ApJ, 412, 443
\bibitem[]{}
Freeman K.C., 1970, ApJ, 160, 811
\bibitem[]{}
Guiderdoni, B., Rocca-Volmerange, B., 1987, A\&A, 186, 1
\bibitem[]{}
Hodge, P.W., 1967, ApJ, 148, 719
\bibitem[]{}
Hoffman, G.L., Salpeter, E.E., Farhat, B., Roos, T.G., Williams, H.L.,
Helou, G., 1996, ApJS, in press
\bibitem[]{}
Impey, C., Bothun, G., 1989, ApJ, 341, 89
\bibitem[]{}
Irwin, M. J., Davies, J. I., Disney, M. J.; Phillipps, S., 1990, MNRAS,
245, 289
\bibitem[]{}
Jobin M., Carignan C., 1990, AJ, 100, 648
\bibitem[]{}
Kent S.M., 1986, AJ, 91, 1301
\bibitem[]{}
Kent S.M., 1987, AJ, 93, 816
\bibitem[]{}
Landolt A.U., 1992, AJ, 104, 340
\bibitem[]{}
Maloney P., 1993, ApJ, 414, 41  
\bibitem[]{}
McGaugh S.S., 1994, ApJ, 426, 135
\bibitem[]{}
McGaugh S.S., 1996, MNRAS, in press
\bibitem[]{}
McGaugh S.S., Bothun G.D., Schombert J.M., 1995, AJ, 110 573
\bibitem[]{}
McGaugh S.S., Schombert J.M., Bothun G.D., 1995, AJ, 109, 2019 
\bibitem[]{}
McGaugh S.S., Bothun G.D., 1994, AJ, 107, 530 (MB94)
\bibitem[]{}
Mo, H. J., McGaugh, S.S., Bothun, G. D., 1994, MNRAS, 267, 129
\bibitem[]{}
Persic M., Salucci P., 1991, ApJ, 368, 60
\bibitem[]{}
Pfenniger D., Combes  F., Martinet L., 1994, A\&A, 285, 79
\bibitem[]{}
Puche D., Carignan C., Bosma A., 1990, AJ, 100, 1468
\bibitem[]{}
Puche D., Carignan C., Wainscoat R.J., 1991, AJ, 101, 447
\bibitem[]{}
Quinn, T., Katz, N., Efstathiou, G., 1995, preprint
\bibitem[]{}
Roberts M.S., Haynes M.P., 1994, A\&AR, 32, 115
\bibitem[]{}
Roelfsema P.R., Allen R.J., 1985, A\&A, 146, 213
\bibitem[]{}
Romanishin, W.,  Krumm, N.,  Salpeter, E.,  Knapp, G.,  Strom, K. M.,
Strom, S. E., 1982, ApJ, 263, 94
\bibitem[]{}
R\" onnback J., 1993, PhD thesis, University of Uppsala
\bibitem[]{}
R\" onnback J., Bergvall N., 1995, preprint
\bibitem[]{}
Schombert J.M., Bothun G.D., Impey C.D., Mundy L.G., 1990, AJ, 100,1523
\bibitem[]{}
Schombert J.M., Bothun G.D., 1988, AJ, 95, 1389
\bibitem[]{}
Schombert J.M., Bothun G.D., Schneider S.E., McGaugh S.S., 1992, AJ,
103, 1107 (SBSM)
\bibitem[]{}
Schwartzenberg J.M., Phillips S., Smith R.M., Couch W.J., Boyle B.J., 1995,
MNRAS, 275, 121
\bibitem[]{}
Sprayberry D., Bernstein G.M., Impey C.D., Bothun G.D., 1995, ApJ, 438, 72
\bibitem[]{}
Sprayberry D., Impey C.D., Bothun G.D., Irwin M., 1995, AJ, 109, 558 
\bibitem[]{}
Turner, J. A.,  Phillipps, S.,  Davies, J. I., Disney, M. J., MNRAS,
1993, 261, 39
\bibitem[]{}
van Albada T.S, Sancisi R., 1986, Phil. Trans. R. Soc. Lond. A, 320, 447
\bibitem[]{}
van der Hulst J.M., Skillman E.D., Smith T.R., Bothun G.D., McGaugh
S.S.,  de Blok W.J.G., 1993, AJ, { 106}, 548 (vdH93)
\bibitem[]{}
van der Hulst J.M., Terlouw J.P., Begeman K.G., Zwitser W., Roelfsema
P.R., 1992, in Worall D.M., Biemesderfer C., Barnes J., ed., 
ASP Conf Series Vol. 25, Astronomical Data Analysis Software and Systems
I, San Francisco, p. 131
\bibitem[]{}
Warmels R.H., 1988, A\&AS, 73, 453 
\bibitem[]{}
Wevers B. M. H. R., van der Kruit P. C., Allen R.J., 1986, A\&AS, 66, 505
\bibitem[]{}
Wilson C.D., 1995, preprint
\bibitem[]{}
Young J.S., Knezek P.M., 1989, ApJ 347, L55
\bibitem[]{}
Zwaan M.A., van der Hulst J.M., de Blok W.J.G., McGaugh S.S., 1995,
MNRAS, 273, L35 (ZHBM)
\end{thebibliography}
\end{document}